\documentclass{SciPost}
\usepackage{amsmath,amssymb,graphicx,bm,color,mathrsfs,verbatim,epstopdf,dcolumn,cancel}

\usepackage{hyperref}
\hypersetup{ 
	colorlinks=true,
}

\usepackage{graphicx}
\usepackage{caption}
\usepackage{subcaption}

\usepackage{listings}
\usepackage{textcomp}
\usepackage[space=true]{accsupp}

\newcommand{\pdfactualhex}[3]{\newcommand{#1}{%
		\BeginAccSupp{method=hex,ActualText=#2}#3\EndAccSupp{}}}

\pdfactualhex{\pdfactualdspace}{2020}{\textperiodcentered\textperiodcentered}
\pdfactualhex{\pdfactualsquote}{27}{'}
\pdfactualhex{\pdfactualbtick}{60}{`}

\definecolor{deepblue}{rgb}{0,0,0.8}
\definecolor{deepred}{rgb}{1.0,0,0}
\definecolor{deepgreen}{rgb}{0,0.7,0}
\definecolor{blueviolet}{RGB}{138,43,226}
\definecolor{darkyellow}{RGB}{204,204,0}
\definecolor{codegray}{rgb}{0.6,0.6,0.6}
\definecolor{weborange}{RGB}{255,165,0}
\definecolor{gold}{RGB}{255,205,0}
\definecolor{codegreen}{rgb}{0,0.6,0}
\definecolor{codepurple}{rgb}{0.58,0,0.82}

\definecolor{backcolour}{rgb}{0.95,0.95,0.92}

\lstdefinestyle{sublime}{
	backgroundcolor=\color{backcolour},   
	commentstyle=\color{deepgreen},
	keywordstyle=\color{deepred},
	numberstyle=\tiny,
	stringstyle=\color{weborange},
	basicstyle=\small\ttfamily, 
	breakatwhitespace=false,         
	breaklines=true,                 
	captionpos=t,                    
	keepspaces=true,                 
	numbers=left,                    
	numbersep=5pt,                  
	showspaces=false,                
	showstringspaces=false,
	showtabs=false,                  
	tabsize=4,
	columns=flexible,
	emptylines=10000,
	literate={'}{\pdfactualsquote}1{`}{\pdfactualbtick}1{\ \ }{\pdfactualdspace}2,
	inputpath=./anc,
	keywords={lambda,xrange,abs,for,return},
}
\usepackage{marvosym,etoolbox}
\usepackage[T1]{fontenc}
\usepackage{graphicx}
\newcommand\0{\scalebox{-1}[1]{0}}
\let\svttfamily\ttfamily

\catcode`0=\active
\def0{\0}
\renewcommand\ttfamily{\svttfamily\catcode`0=\active }
\renewcommand\texttt{\bgroup\ttfamily\texttthelp}
\def\texttthelp#1{#1\egroup}
\catcode`0=12 %

\graphicspath{{figs/}}

\usepackage{background}
\usepackage{geometry}

\definecolor{textcolor}{HTML}{0A75A8}
\newcommand\Text{ \emph{to report a bug pls visit https://github.com/weinbe58/QuSpin/issues} }

\SetBgColor{textcolor}
\SetBgOpacity{0.5}
\SetBgAngle{-90}
\SetBgPosition{current page.center}
\SetBgVshift{0.35\textwidth}
\SetBgScale{1.8}
\SetBgContents{\sffamily\Text}

\newcommand{\JWcode}{example4.py}
\newcommand{\SSHcode}{example5.py}
\newcommand{\MBLcode}{example6.py}
\newcommand{\BHLcode}{example7.py}
\newcommand{\GPcode}{example8.py}
\newcommand{\Spincode}{example9.py}
\newcommand{\BFMcode}{example10.py}

\begin{document}
\begin{center}{\Large \textbf{
QuSpin: a Python Package for Dynamics and Exact Diagonalisation of Quantum Many Body Systems.\\
\large Part II: bosons, fermions and higher spins
}}\end{center}

\begin{center}
Phillip Weinberg\textsuperscript{1*} and Marin Bukov\textsuperscript{2}
\end{center}

\begin{center}
$^1$Department of Physics, Boston University, \\
590 Commonwealth Ave., Boston, MA 02215, USA
\\
* weinbe58@bu.edu
\end{center}

\begin{center}
$^2$Department of Physics, University of California Berkeley, \\
CA 94720, USA
\\
\end{center}

\begin{center}
\today
\end{center}


\section*{Abstract}
{\bf 
We present a major update to QuSpin, \emph{SciPostPhys.2.1.003} -- an open-source Python package for exact diagonalization and quantum dynamics of arbitrary boson, fermion and spin many-body systems, supporting the use of various (user-defined) symmetries in one and higher dimension and (imaginary) time evolution following a user-specified driving protocol. We explain how to use the new features of QuSpin using seven detailed examples of various complexity: (i) the transverse-field Ising chain and the Jordan-Wigner transformation, (ii) free particle systems: the Su-Schrieffer-Heeger (SSH) model, (iii) the many-body localized 1D Fermi-Hubbard model, (iv) the Bose-Hubbard model in a ladder geometry, (v) nonlinear (imaginary) time evolution and the Gross-Pitaevskii equation on a 1D lattice, (vi) integrability breaking and thermalizing dynamics in the translationally-invariant 2D transverse-field Ising model, and (vii) out-of-equilibrium Bose-Fermi mixtures. This easily accessible and user-friendly package can serve various purposes, including educational and cutting-edge experimental and theoretical research. The complete package documentation is available under \href{http://weinbe58.github.io/QuSpin/index.html}{http://weinbe58.github.io/QuSpin/index.html}.
}

\vspace{10pt}
\noindent\rule{\textwidth}{1pt}
\tableofcontents\thispagestyle{fancy}
\noindent\rule{\textwidth}{1pt}
\vspace{10pt}

\section{What Problems can I Study with QuSpin?}
\label{sec:intro}

Understanding the physics of many-body quantum condensed matter systems often involves a great deal of numerical simulations, be it to gain intuition about the complicated problem of interest, or because they do not admit an analytical solution which can be expressed in a closed form. This motivates the development of open-source packages~\cite{alet05,albuquerque2007,bauer11,dolfi14,ITensor,TNT,johansson2012,johansson2013,wright_13,kramer_17,gegg_17,young_17,al_17}, the purpose of which is to facilitate the study of condensed matter systems, without the need to implement the inner workings of complicated numerical methods which required years to understand and fully develop. Here, we report on a major upgrade to QuSpin~\cite{weinberg_17_quspin} -- a Python library for exact diagonalisation (ED) and simulation of the dynamics of arbitrary quantum many-body systems. 

Although ED methods are vastly outperformed by more sophisticated numerical techniques in the study of equilibrium problems, such as quantum Monte Carlo~\cite{pollet_12,foulkes_01,acioli_97}, matrix product states based density matrix renormalisation group~\cite{daley_04,schollwock_05,schollwock_11}, and dynamical mean-field theory~\cite{georges_96,kotliar_06,aoki_14}, as of present date ED remains essential for certain dynamical non-equilibrium problems. The reason for this often times relies on the fact that the underlying physics of these problems cannot be explained without taking into consideration the contribution from high-energy states excited during the nonequilibrium process. Some prominent examples of such problems include the study of the many-body localisation (MBL) transition~\cite{altman2015universal,nandkishore_15,abanin_17,thomson_18,rubio2018probing}, the Eigenstate Thermalisation hypothesis~\cite{TD_review}, ergodicity breaking, thermalization and scrambling~\cite{nation2019ergodicity,duval2019quantum,bentsen2019treelike}, quantum quench dynamics~\cite{polkovnikov_11}, periodically-driven systems~\cite{goldman_14,bukov_review,eckardt_17,claeys2017breaking,claeys2017spin,vajna2017replica,weinberg_FAPT,howell2018frequency,vogl2019flow}, non-demolition measurements in many-body systems~\cite{yang2019quantum}, long-range quantum coherence~\cite{mzaouali2019long}, dynamics-induced instabilities~\cite{niu_01,creffield_09,bukov_12,citro2015dynamical,bukov_17RL,lellouch_17,naeger2018parametric,boulier2018parametric}, adiabatic and counter-diabatic state preparation~\cite{kolodrubetz_17,delcampo_13,sels_16,bukov_GSL,claeys2019floquet}, dynamical~\cite{heyl2018dynamical,de2018stochastic} and topological~\cite{zache2019dynamical} phase transitions applications of Machine Learning to (non-equilibrium) physics~\cite{mehta2019high,bukov_17RL,dunjko_17,carleo2019machine,bukov2018reinforcement,torlai2019integrating,fabiani2019investigating}, optimal control~\cite{glaser_15,bukov_17symmbreak,day2018glassy}, to benchmark results in sophisticated algorithms, such as quantum annealing~\cite{patil2019obstacles} and Quantum Monte Carlo~\cite{kubiczek2019exact}, and many more. Besides dynamical studies, ED methods are currently heavily used to compute the full (or low-energy) spectrum of frustrated Hamiltonians, where they compete with a high-precision tensor network approach. Last but not least, virtually all new numerical techniques are frequently benchmarked against ED.

It is, thus, arguably useful to have a toolbox available at one's disposal which allows one to quickly simulate and study these and related nonequilibrium problems. As such, QuSpin offers easy access to performing numerical simulations, which can facilitate the development and inspiration of new ideas and the discovery of novel phenomena, eliminating the cost of spending time to develop a reliable code. Besides theorists, the new version of QuSpin will hopefully even prove valuable to experimentalists working on problems containing dynamical setups, as it can help students and researchers focus on perfecting the experiment, rather than worry about writing the supporting simulation code. Last but not least, with the computational processing power growing higher than ever before, the role played by simulations for theoretical research becomes increasingly more important too. It can, therefore, be expected that in the near future quantum simulations become an integral part of the standard physics university curriculum, and having easily accessible toolboxes, such as QuSpin, is one of the requisites for this anticipated change.

\section{How do I Use the New Features of QuSpin?}
\label{sec:examples}

New in QuSpin 3.0, we have added the following features and toolboxes:
\begin{itemize}
	\item[(i)] support for fermion, boson and higher-spin Hamiltonians with \texttt{basis.ent\_entropy} and \texttt{basis.partial\_trace} routine to calculate entanglement entropy, reduced density matrices, and entanglement spectrum.
	\item[(ii)] \texttt{general} basis constructor classes for user-defined symmetries which allow the implementation of higher-dimensional lattice structures.
	\item[(iii)] \texttt{block\_ops} class and \texttt{block\_diag\_hamiltonian} function to automatically handle splitting the evolution over various symmetry sectors of the Hilbert space.
	\item[(iv)] user-customisable \texttt{evolve} routine to handle user-specified linear and non-linear equations of motion.
	\item[(v)] \texttt{quantum\_operator} class to define parameter-dependent Hamiltonians.
	\item[(vi)] \texttt{quantum\_LinearOperator} class which applies the operator "on the fly" which saves vast amounts of memory at the cost of computational time. 
\end{itemize}

\noindent Before we carry on, we refer the interested reader to \href{http://weinbe58.github.io/QuSpin/Examples.html}{Examples 0-3} from the original QuSpin paper~\cite{weinberg_17_quspin}. The examples below focus predominantly on the newly introduced features, and are thus to be considered complementary. We emphasize that, since they serve the purpose of explaining how to use QuSpin, for the sake of brevity we shall not discuss the interesting physics related to the interpretation of the results.\\

\emph{Installing QuSpin is quick and efficient; just follow the steps outlined in App.~\ref{app:install}.}

\subsection{The Spectrum of the Transverse Field Ising Model and the Jordan-Wigner Transformation}
\label{subsec:JW}

This example shows how to
\begin{itemize}
	\item construct fermionic hopping, $p$-wave pairing and on-site potential terms, and spin$-1/2$ interactions and transverse fields,
	\item implement periodic and anti-periodic boundary conditions,
	\item use particle conservation modulo $2$, and spin inversion,
	\item handle the default built-in particle conservation and symmetry checks,
	\item obtain the spectrum of a QuSpin Hamiltonian.
\end{itemize}

\noindent\emph{Physics Setup---}The transverse field Ising (TFI) chain is paradigmatic in our understanding of quantum phase transitions, since it represents an exactly solvable model~\cite{sachdev_book}. The Hamiltonian is given by
\begin{equation}
H=\sum_{j=0}^{L-1}-J\sigma^z_{j+1}\sigma^z_j - h\sigma^x_j,
\label{eq:TFIM}
\end{equation} 
where the nearest-neighbour (nn) spin interaction is $J$, $h$ denotes the transverse field, and $\sigma^\alpha_j$ are the Pauli spin-$1/2$ matrices. We use periodic boundary conditions and label the $L$ lattice sites $0,\dots,L-1$ to conform with Python's convention. This model has gapped, fermionic elementary excitations, and exhibits a phase transition from an antiferromagnet to a paramagnet at $\left(h/J\right)_c=1$. The Hamiltonian possesses the following symmetries: parity (reflection w.r.t.~the centre of the chain), spin inversion, and (many-body) momentum conservation.

In one dimension, the TFI Hamiltonian can be mapped to spinless $p$-wave superconducting fermions via the Jordan-Wigner (JW) transformation~\cite{sachdev_book,dziarmaga_10,pfeuty_79}:
\begin{equation}
c_i=\frac{\sigma^x_i-i\sigma^y_i}{2}\prod_{j<i}\sigma^z_j,\qquad c^\dagger_i=\frac{\sigma^x_i+i\sigma^y_i}{2}\prod_{j<i}\sigma^z_j,
\label{eq:JW_transf}
\end{equation} 
where the fermionic operators satisfy $\{c_i,c^\dagger_j\}=\delta_{ij}$. The Hamiltonian is readily shown to take the form
\begin{equation}
H=\sum_{j=0}^{L-1}J\left(-c^\dagger_jc_{j+1} + c_jc^\dagger_{j+1} \right) +J\left( -c^\dagger_jc^\dagger_{j+1} + c_jc_{j+1}\right) + 2h\left(n_j-\frac{1}{2}\right).
\label{eq:TFIM_fermion}
\end{equation}
In the fermionic representation, the spin $zz$-interaction maps to nn hopping and a $p$-wave pairing term with coupling constant $J$, while the transverse field translates to an on-site potential shift of magnitude $h$. In view of the implementation of the model using QuSpin, we have ordered the terms such that the site index is growing to the right, which comes at the cost of a few negative signs due to the fermion statistics. We emphasize that this ordering is not required by QuSpin, but it is merely our choice to use it here for the sake of consistency. The fermion Hamiltonian posses the symmetries: particle conservation modulo $2$, parity and (many-body) ``momentum'' conservation.

Here, we are interested in studying the spectrum of the TFI model in both the spin and fermion representations. However, if one na\"ively carries out the JW transformation, and computes the spectra of Eqs.~\eqref{eq:TFIM} and~\eqref{eq:TFIM_fermion}, one might be surprised that they do not match exactly. The reason lies in the form of the boundary condition required to make the JW mapping exact -- a subtle issue often left aside in favour of discussing the interesting physics of the TFI model itself. 

We recall that the starting point is the periodic boundary condition imposed on the spin Hamiltonian in Eq.~\eqref{eq:TFIM}. Due to the symmetries of the spin Hamiltonian~\eqref{eq:TFIM}, we can define the JW transformation on every symmetry sector separately. To make the JW mapping exact, we supplement Eq.~\eqref{eq:JW_transf} with the following boundary conditions: (i) the negative spin-inversion symmetry sector maps to the fermion Hamiltonian~\eqref{eq:TFIM_fermion} with \emph{periodic} boundary conditions (PBC) and \emph{odd} total number of fermions; (ii) the positive spin-inversion symmetry sector maps to the fermion Hamiltonian~\eqref{eq:TFIM_fermion} with \emph{anti-periodic} boundary conditions (APBC) and \emph{even} total number of fermions. Anti-periodic boundary conditions differ from PBC by a negative sign attached to all coupling constants that cross a single, fixed lattice bond (the bond itself is arbitrary as all bonds are equal for PBC). APBC and PBC are special cases of the more general twisted boundary conditions where, instead of a negative sign, one attaches an arbitrary phase factor.

\begin{figure}[t!]
	\centering
	\begin{subfigure}[b]{0.496\textwidth}
		\includegraphics[width=\textwidth]{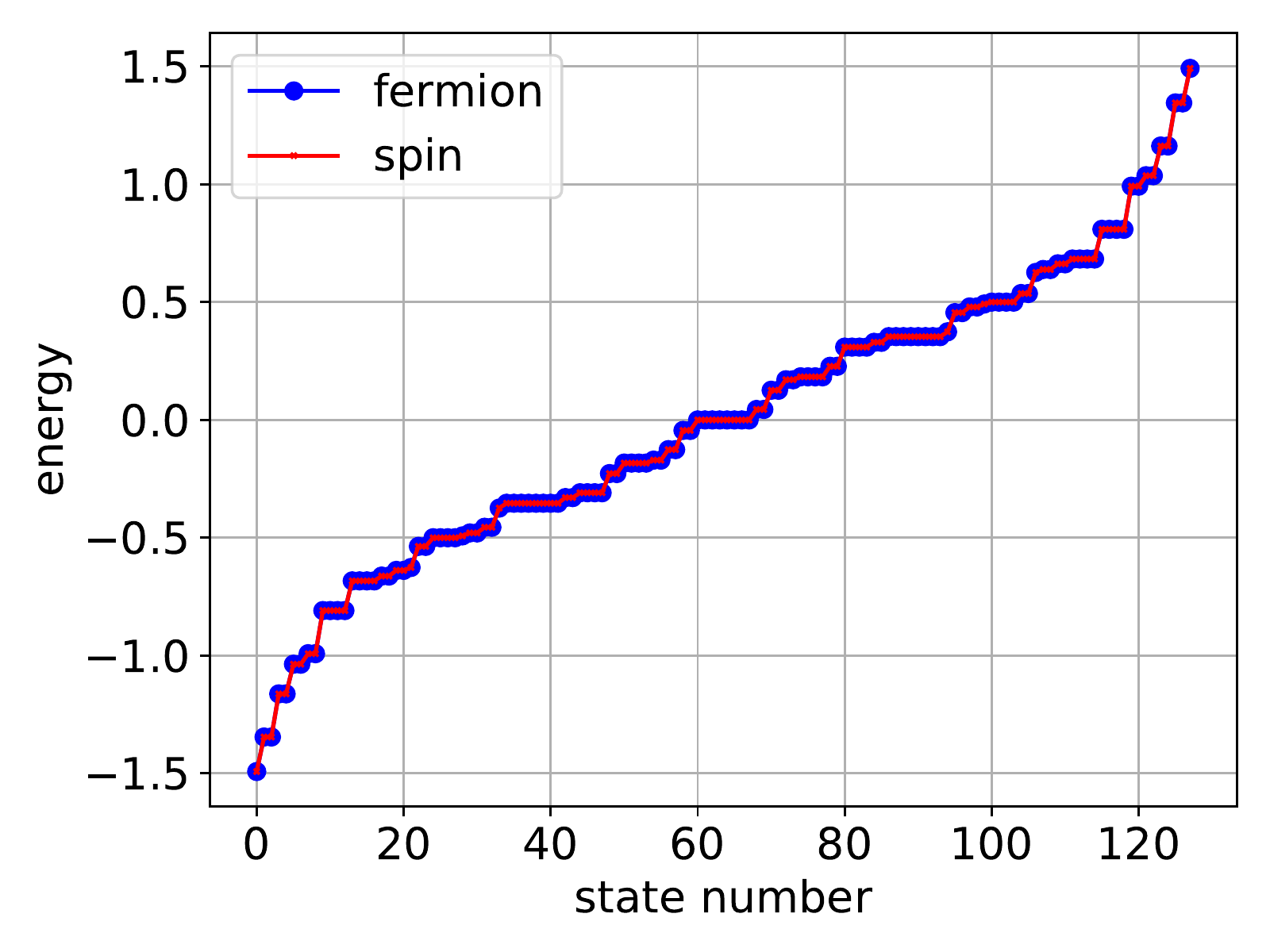}
		\caption{negative spin inversion/PBC sector}
	\end{subfigure}
	\begin{subfigure}[b]{0.496\textwidth}
		\includegraphics[width=\textwidth]{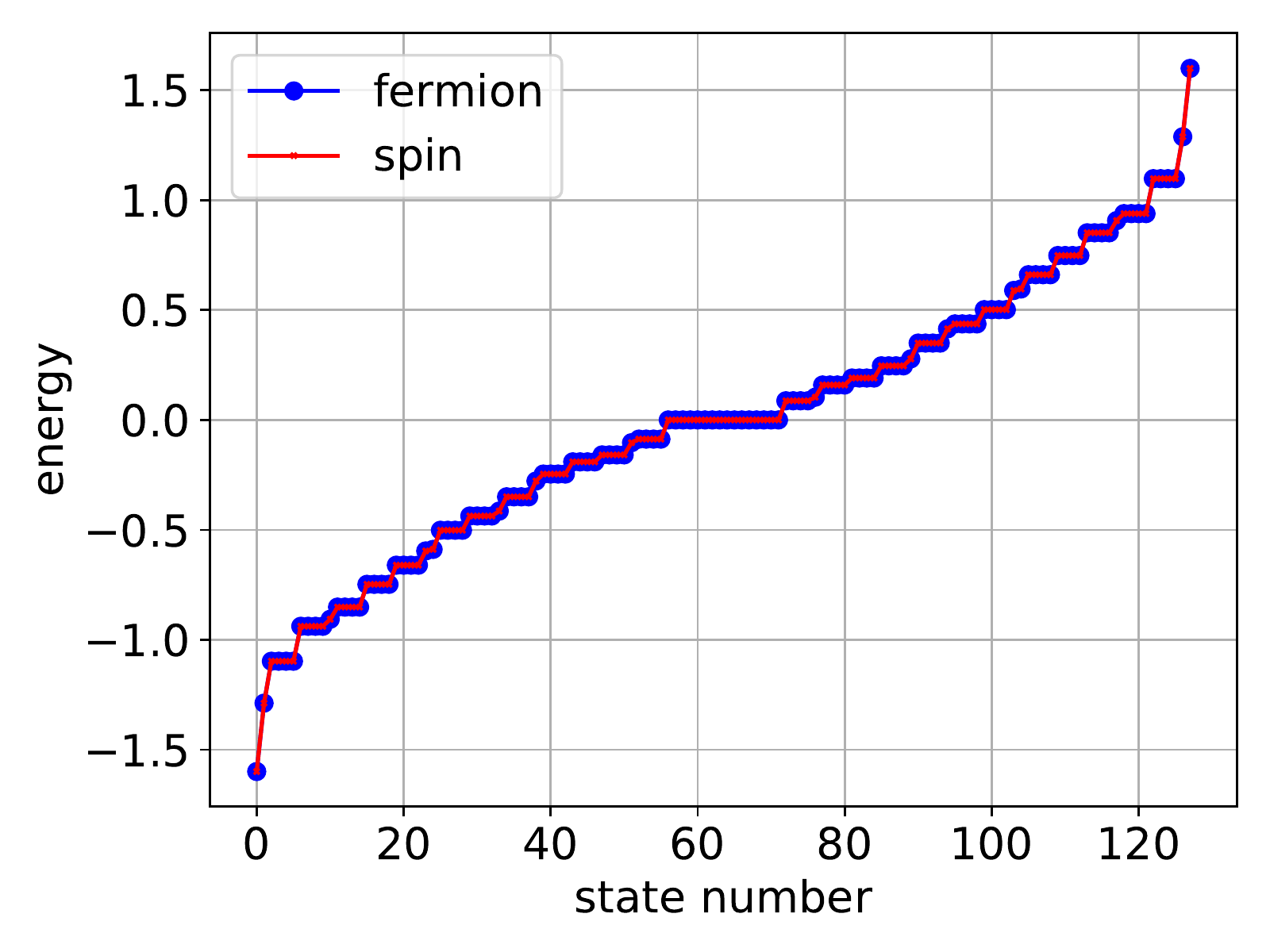}
		\caption{positive spin inversion/APBC sector}
	\end{subfigure}
	\caption{\label{fig:JW} Comparison of the spectra [in units of $J$] of the spin~\eqref{eq:TFIM} and fermion~\eqref{eq:TFIM_fermion} representation of the transverse field Ising Hamiltonian. The degeneracy in the spectrum is due to the remaining reflection and translation symmetries which could easily be taken into account (see text). The parameters are $J=1.0$, $h=\sqrt{2}$, and $L=8$.}  
\end{figure}

In the following, we show how to compute the spectra of the Hamiltonians in Eqs.~\eqref{eq:TFIM} and~\eqref{eq:TFIM_fermion} with the correct boundary conditions using QuSpin. Figure~\ref{fig:JW} shows that they match exactly in both the PBC and APBC cases discussed above, as predicted by theory.

\noindent\emph{Code Analysis---}We begin by loading the QuSpin operator and basis constructors, as well as some standard Python libraries. 
\lstinputlisting[firstline=1, lastline=4, firstnumber=1]{\JWcode}
First, we define the models parameters.
\lstinputlisting[firstline=6, lastline=9, firstnumber=6]{\JWcode}
We have to consider two cases when computing the spectrum, as discussed in the theory section above. In one case, the fermionic system has PBC and the particle number sector is odd, while the spins are constrained to the negative spin inversion symmetry sector, while in the second -- the fermion model has APBC with even particle number sector, and the spin model is considered in the positive spin inversion sector. To this end, we introduce the variables \texttt{zblock} $\in\{\pm 1\}$ and \texttt{PBC} $\in\{\pm 1\}$, where \texttt{PBC} $=-1$ denotes APBC. Note that the only meaningful combinations are $($\texttt{zblock}, \texttt{PBC}$)=(-1,1),(1,-1)$, which we loop over:
\lstinputlisting[firstline=11, lastline=12, firstnumber=11]{\JWcode}
Within this loop, the code is divided in two independent parts: first, we compute the spectrum of the TFI system, and then -- that of the equivalent fermionic model. Let us discuss the spins. 

In QuSpin, operators are stored as sparse lists. These lists contain two parts: (i) the lattice sites on which the operator acts together with the coupling strength, which we call a \emph{site-coupling} list, and (ii) the types of the operators involved, i.e.~the \emph{operator-string}. For example, the operator $\mathcal{O}=g\sum_{j=0}^{L-1}\sigma^\mu_j$ can be uniquely represented by the site-coupling list \texttt{[[g,0],[g,1],$\dots$,[g,L-1]]}, and the information that it is the Pauli matrix $\mu$. The components lists are nothing but the pairs of the field strength and the site index \texttt{[g,j]}. It is straightforward to generalise this to non-uniform fields \texttt{g$\;\mapsto\;$g[j]}. Similarly, any two-body operator $\mathcal{O}=J_{zz}\sum_{j=0}^{L-1}\sigma^\mu_j\sigma^\nu_{j+1}$ can be fully represented by the two sites it acts on, and its coupling strength: \texttt{[J,j,j+1]}. We then stack up these elementary lists together into the full site-coupling list: \texttt{[[J,0,1],[J,1,2],$\dots$,[J,L-2,L-1],[J,L-1,0]]}. 
\lstinputlisting[firstline=14, lastline=17, firstnumber=14]{\JWcode}
Notice the way we defined the periodic boundary condition for the spin-spin interaction using the modulo operator \texttt{\%}, which effectively puts a coupling between sites $L-1$ and $0$. We mention in passing that the above procedure generalises so one can define any multi-body local or nonlocal operator using QuSpin. 

In order to specify the types of the on-site single-particle operators, we use operator strings. For instance, the transverse field operator $\mathcal{O}=g\sum_{j=0}^{L-1}\sigma^x_j$ becomes \texttt{['x',h\_field]}, while the two-body $zz$-interaction is \texttt{['zz',J\_zz]}. It is important to notice that the order of the letters in the operator string corresponds to the order the operators are listed in the site-coupling lists. Putting everything into one final list yields the \texttt{static} list for the spin model:  
\lstinputlisting[firstline=18, lastline=19, firstnumber=18]{\JWcode}
In QuSpin, the user can define both static and dynamic operators. Since this example does not contain any time-dependent terms, we postpone the explanation of how to use dynamic lists to Sec.~\ref{subsec:GP_dynamics}, and use an empty list instead.
\lstinputlisting[firstline=20, lastline=20, firstnumber=20]{\JWcode}
The last step before we can construct the Hamiltonian is to build the basis for it. This is done using the \texttt{basis} constructors. For spin systems, we use \texttt{spin\_basis\_1d} which supports the operator strings \texttt{'z','+','-','I'}, and for spin-$1/2$ additionally \texttt{'x','y'}. The first and required argument is the number of sites $L$. Optional arguments are used to parse symmetry sectors. For instance, if we want to construct an operator in the spin-inversion block with quantum number $+1$, we can conveniently do this using the flag \texttt{zblock=1}.
\lstinputlisting[firstline=21, lastline=22, firstnumber=21]{\JWcode}
Having specified the static and dynamic lists, as well as the basis, building up the Hamiltonian is a one-liner, using the \texttt{hamiltonian} constructor. The required arguments in order of appearance are the \texttt{static} and \texttt{dynamic} lists, respectively. Optional arguments include the \texttt{basis}, and the precision or data type \texttt{dtype}. If no basis is passed, the constructor uses \texttt{spin\_basis\_1d} by default. The default data type is \texttt{np.complex128}.
\lstinputlisting[firstline=23, lastline=24, firstnumber=23]{\JWcode} 
The Hamiltonian is stored as a Scipy sparse matrix for efficiency. It can be converted to a dense array for a more convenient inspection using the attribute \texttt{H.toarray()}. To calculate its spectrum, we use the attribute \texttt{H.eigenvalsh()}, which returns all eigenvalues. Other attributes for diagonalisation were discussed \href{http://weinbe58.github.io/QuSpin/examples/example0.html#example0-label}{here}, c.f.~Ref.~\cite{weinberg_17_quspin}.
\lstinputlisting[firstline=25, lastline=26, firstnumber=25]{\JWcode} 

Let us now move to the second part of the loop which defines the fermionic $p$-wave superconductor. We start by defining the site-coupling list for the local potential
\lstinputlisting[firstline=28, lastline=30, firstnumber=28]{\JWcode}
Let us focus on the case of periodic boundary conditions \texttt{PBC=1} first. 
\lstinputlisting[firstline=31, lastline=31, firstnumber=31]{\JWcode}
In the fermion model, we have two types of two-body terms: hopping terms $c^\dagger_{i}c_{i+1} - c_{i}c^\dagger_{i+1}$, and pairing terms $c^\dagger_{i}c^\dagger_{i+1} - c_{i}c_{i+1}$. While QuSpin allows any ordering of the operators in the static and dynamic lists, for the sake of consistency we set a convention: the site indices grow to the right. To take into account the opposite signs resulting from the fermion statistics, we have to code the site-coupling lists for all four terms separately. These two-body terms are analogous to the spin-spin interaction above:
\lstinputlisting[firstline=32, lastline=36, firstnumber=32]{\JWcode}
To construct a fermionic operator, we make use of the basis constructor \texttt{spinless\_fermion\_basis\_1d}. Once again, we pass the number of sites \texttt{L}. As we explained in the analysis above, we need to consider all odd particle number sectors in the case of PBC. This is done by specifying the particle number sector \texttt{Nf}. 
\lstinputlisting[firstline=37, lastline=38, firstnumber=37]{\JWcode}

In the case of APBC, we first construct all two-body site-coupling lists as if the boundaries were open, and supplement the APBC links with negative coupling strength in the end:
\lstinputlisting[firstline=39, lastline=49, firstnumber=39]{\JWcode}
The construction of the basis is the same, except that this time we need all even particle number sectors:
\lstinputlisting[firstline=50, lastline=51, firstnumber=50]{\JWcode}

As before, we need to specify the type of operators that go in the fermion Hamiltonian using operator string lists. The \texttt{spinless\_fermion\_basis\_1d} class accepts the following strings \texttt{'+','-','n','I'}, and additionally the particle-hole symmetrised density operator \texttt{'z'} $=n-1/2$. The static and dynamic lists read as
\lstinputlisting[firstline=52, lastline=54, firstnumber=52]{\JWcode}
Constructing and diagonalising the fermion Hamiltonian is the same as for the spin-$1/2$ system. Note that one can disable the automatic built-in checks for particle conservation \texttt{check\_pcon=False} and all other symmetries \texttt{check\_symm=False} if one wishes to suppress the checks. 
\lstinputlisting[firstline=55, lastline=59, firstnumber=55]{\JWcode}

The complete code including the lines that produce Fig.~\ref{fig:JW} is available under:\\

\href{http://weinbe58.github.io/QuSpin/examples/example4.html}{http://weinbe58.github.io/QuSpin/examples/example4.html}\\

\subsection{Free Particle Systems: the Fermionic Su-Schrieffer-Heeger (SSH) Chain}

\label{subsec:SSH_model}

This example shows how to
\begin{itemize}
	\item construct free-particle Hamiltonians in real space,
	\item implement translation invariance with a two-site unit cell and construct the single-particle momentum-space block-diagonal Hamiltonian using the \texttt{block\_diag\_hamiltonian} tool,
	\item compute non-equal time correlation functions,
	\item time-evolve multiple quantum states simultaneously.
\end{itemize}

\noindent\emph{Physics Setup---}The Su-Schrieffer-Heeger (SSH) model of a free-particle on a dimerised chain was first introduced in the seventies to study polyacetylene~\cite{SSH_1,SSH_2,SSH_2_E}. Today, this model is paradigmatically used in one spatial dimension to introduce the concepts of Berry phase, topology, edge states, etc~\cite{asboth2016short}. The Hamiltonian is given by
\begin{eqnarray}
H = \sum_{j=0}^{L-1} -(J+(-1)^j\delta J)\left(c_jc^\dagger_{j+1} - c^\dagger_{j}c_{j+1}\right) + \Delta(-1)^jn_j, 
\label{eq:SSH}
\end{eqnarray}
where $\{c_i,c^\dagger_j\}=\delta _{ij}$ obey fermionic commutation relations. The uniform part of the hopping matrix element is $J$, the bond dimerisation is defined by $\delta J$, and $\Delta$ is the staggered potential. We work with periodic boundary conditions.

Below, we show how one can use QuSpin to study the physics of free fermions in the SSH chain. One way of doing this would be to work in the many-body (Fock space) basis, see Sec.~\ref{subsec:Fermion_MBL}. However, whenever the particles are non-interacting, the exponential scaling of the Hilbert space dimension with the number of lattice sites imposes an artificial limitation on the system sizes one can do. Luckily, with no interactions present, the many-body wave functions factorise in a product of single-particle states. Hence, it is possible to study the behaviour of many free bosons and fermions by simulating the physics of a single particle, and populating the states according to bosonic or fermionic statistics, respectively.

\begin{figure}[t!]
	\centering
	\begin{subfigure}[b]{0.496\textwidth}
		\includegraphics[width=\textwidth]{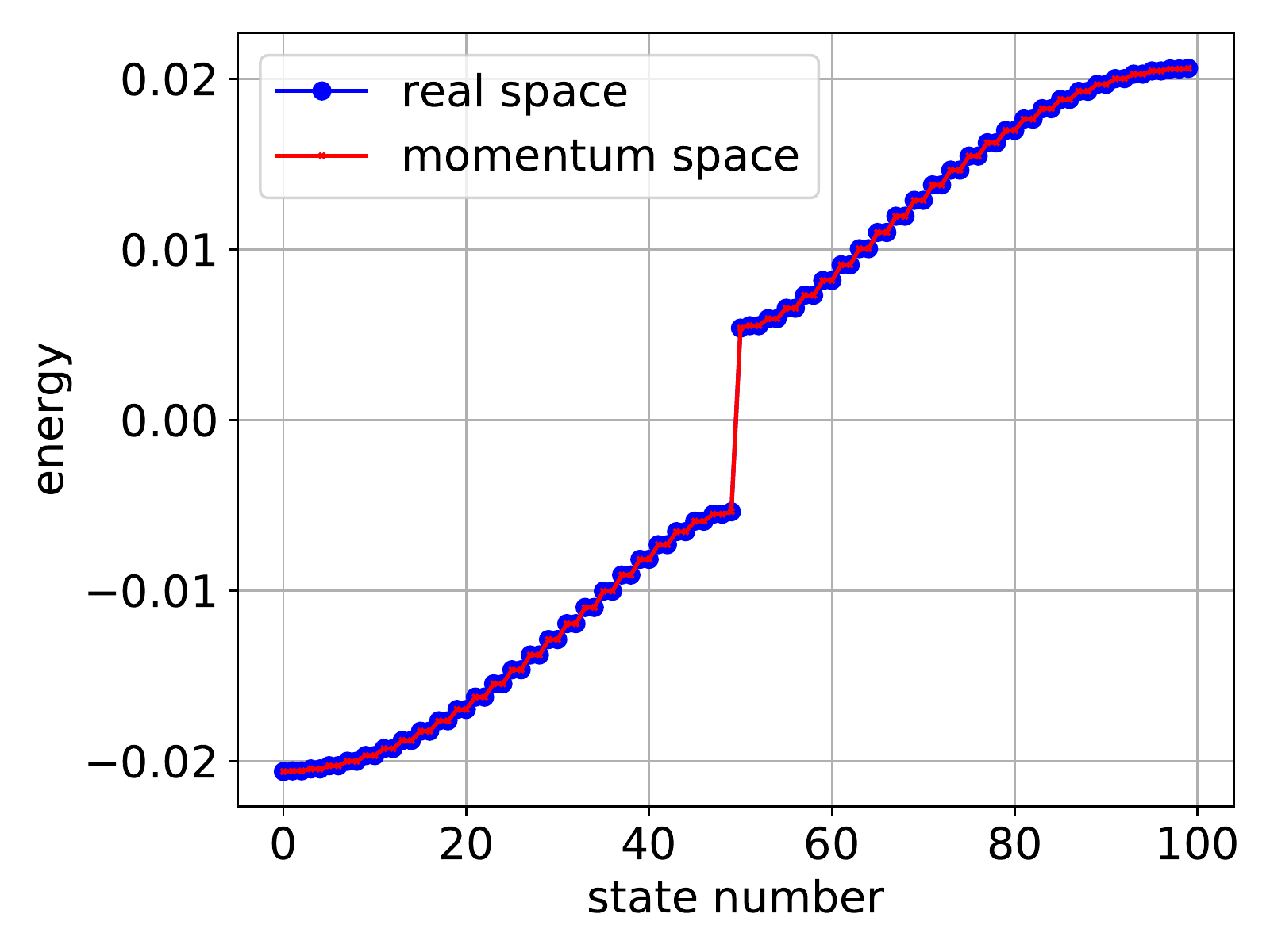}
	\end{subfigure}
	\begin{subfigure}[b]{0.496\textwidth}
		\includegraphics[width=\textwidth]{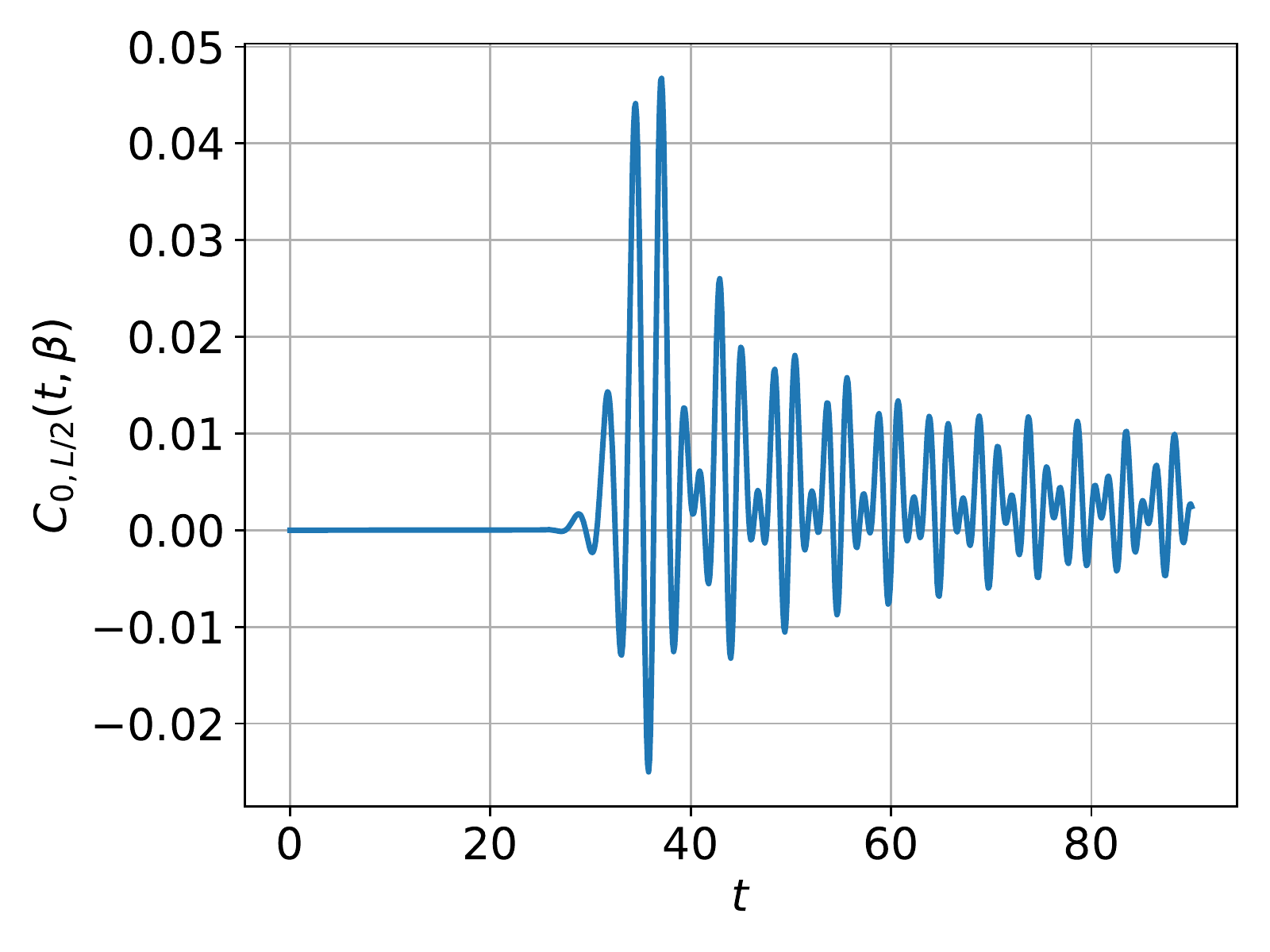}
	\end{subfigure}
	\caption{\label{fig:SSH} (a) the spectrum [in units of $J$] of the SSH Hamiltonian in real and momentum space. (b) The non-equal time correlator $C_{i=0,j=L/2}(t)$, cf.~Eq.~\eqref{eq:SSH_corr} as a function of time. The parameters are $\delta J/J=0.1$, $\Delta/J=0.5$, $J\beta=100.0$ and $L=100$.}  
\end{figure}

Making use of translation invariance, a straightforward Fourier transformation to momentum space, $a_k = \sqrt{2/L}\sum_{j:\mathrm{even}}^{L-1} \mathrm e^{-i k j}c_{j}$ and $b_k = \sqrt{2/L}\sum_{j:\mathrm{odd}}^{L-1}\mathrm e^{-i k j} c_{j}$, casts the SSH Hamiltonian in the following form
\begin{equation}
H \!=\! \sum_{k\in\mathrm{BZ'}} (a^\dagger_k,b^\dagger_k)
\left(\begin{array}{cc}
\Delta & -(J+\delta J)\mathrm e^{-i k} - (J-\delta J)\mathrm e^{+i k} \\
-(J+\delta J)\mathrm e^{+i k} - (J-\delta J)\mathrm e^{-i k} & -\Delta
\end{array}
\right)
\left(\! \begin{array}{c}
a_k\\
b_k
\end{array}
\!\right),
\end{equation}
where the reduced Brillouin zone is defined as $\mathrm{BZ'}=[-\pi/2,\pi/2)$. We thus see that the Hamiltonian reduces further to a set of independent $2\times 2$ matrices. The spectrum of the SSH model is gapped and, thus, has two bands, see Fig.~\ref{fig:SSH}a.

Since we are dealing with free fermions, the ground state is the Fermi sea, $|\mathrm{FS}\rangle$, defined by filling up the lowest band completely. We are interested in measuring the real-space non-equal time density autocorrelation function
\begin{equation}
\label{eq:SSH_corr}
C_{ij}(t) = \langle \mathrm{FS}|n_i(t)n_j(0)|\mathrm{FS}\rangle = \langle \mathrm{FS}(t)|n_i(0)\underbrace{U(t,0)n_j(0)|\mathrm{FS}\rangle}_{|\mathrm{nFS}(t)\rangle}.
\end{equation}
To evaluate the correlator numerically, we shall use the right-hand side of this equation.

As we are studying free particles, it is enough to work with the single-particle states. For instance, the Fermi sea can be obtained as $|\mathrm{FS}\rangle = \prod_{k\leq k_F}c^\dagger_k|0\rangle$. Denoting the on-site density operator by $n_{i}$, one can cast the correlator in momentum space in the following form:
\begin{equation}
C_{ij}(t) = \sum_{k\leq k_F}\langle k|n_i(t)\hat n_j(0)|k\rangle.
\end{equation}
If we want to consider finite-temperature $\beta^{-1}$, the above formula generalises to
\begin{equation}
C_{ij}(t,\beta) = \sum_{k\in\mathrm{BZ'}}n_\mathrm{FD}(k,\beta)\langle k|n_i(t) n_j(0)|k\rangle,
\end{equation} 
where $n_\mathrm{FD}(k,\beta) = 1/(\exp(\beta E_k)+1)$ is the Fermi-Dirac distribution at temperature $\beta^{-1}$, with $E_k$ the SSH dispersion. Figure~\ref{fig:SSH}b) shows the time evolution of $C_{ij}(t,\beta)$ for two sites, separated by the maximal distance on the ring: $L/2$. 

\noindent\emph{Code Analysis.---}Let us explain how one can do all this quickly and efficiently using QuSpin. As always, we start by loading the required packages and libraries.
\lstinputlisting[firstline=1, lastline=9]{\SSHcode}
After that, we define the model parameters
\lstinputlisting[firstline=10, lastline=15]{\SSHcode}
In the following, we construct the fermionic SSH Hamiltonian first in real space. We then show how one can also construct it in momentum space where, provided we use periodic boundary conditions, it appears bock-diagonal. Let us define the fermionic site-coupling lists. Once again, we emphasise that fermion systems require special care in defining the hopping terms: Eq.~\eqref{eq:SSH} is conveniently cast in the form where all site indices on the operators grow to the right, and all signs due to the fermion statistics are explicitly spelt out.  
\lstinputlisting[firstline=16, lastline=20]{\SSHcode}
To define the \texttt{static} list we assign the corresponding operator strings to he site-coupling lists. Since our problem does not feature any explicit time dependence, we leave the \texttt{dynamic} list empty.
\lstinputlisting[firstline=21, lastline=23]{\SSHcode}
Setting up the fermion basis with the help of the constructor \texttt{spinless\_fermion\_basis\_1d} proceeds as smoothly as in Sec.~\ref{subsec:JW}. Notice a cheap trick: by specifying a total of \texttt{Nf=1} fermion in the lattice, QuSpin actually allows to define single-particle models, as a special case of the more general many-body Hamiltonians. Compared to many body models, however, due to the exponentially reduced Hilbert space size, this allows us to scale up the system size \texttt{L}. 
\lstinputlisting[firstline=24, lastline=25]{\SSHcode}
We then build the real-valued SSH Hamiltonian in real space by passing the static and dynamic lists, as well as the basis and the data type. After that, we and diagonalise it, storing all eigenenergies and eigenstates.
\lstinputlisting[firstline=26, lastline=29]{\SSHcode}

For translation invariant single-particle models, however, the user might prefer to use momentum space, where the Hamiltonian becomes block diagonal, as we showed in the theory section above. This can be achieved using QuSpin's \texttt{block\_tools}. The idea behind this tool is simple: the main purpose is to create the full Hamiltonian in block-diagonal form, where the blocks correspond to pre-defined quantum numbers. In our case, we would like to use momentum or \texttt{kblock}'s. Note that the unit cell in the SSH model contains two sites, which we encode using the variable \texttt{a=2}. Thus, we can create a list of dictionaries \texttt{blocks}, each element of which defines a single symmetry block. If we combine all blocks, we exhaust the full Hilbert space. All other basis arguments, such as the system size, we store in the variable \texttt{basis\_args}. We invite the interested user to check the package \href{http://weinbe58.github.io/QuSpin/index.html}{documentation} for additional optional arguments and functionality of \texttt{block\_tools}, cf.~App.~\ref{app:doc}. We mention in passing that this procedure is independent of the symmetry, and can be done using all symmetries supported by QuSpin, not only translation.
\lstinputlisting[firstline=30, lastline=33]{\SSHcode}
To create the block-diagonal Hamiltonian, we invoke the \texttt{block\_diag\_hamiltonian} method. It takes both required and optional arguments, and returns the transformation which block-diagonalises the Hamiltonian (in our case the Fourier transform) and the block-diagonal Hamiltonian object. Required arguments, in order of appearance, are the \texttt{blocks}, the \texttt{static} and \texttt{dynamic} lists, the \texttt{basis} constructor, \texttt{basis\_args}, and the data type. Since we anticipate the matrix elements of the momenum-space Hamiltonian to contain the Fourier factors $\exp(-ik)$, we know to choose a complex data type. \texttt{block\_diag\_hamiltonian} also accepts some optional arguments, such as the flags for disabling the automatic built-in symmetry checks. More about this function can be found in the \href{http://weinbe58.github.io/QuSpin/index.html}{documentation}, cf.~App.~\ref{app:doc}.
\lstinputlisting[firstline=34, lastline=36]{\SSHcode}
We can then use all functions and methods of the \texttt{hamiltonian} class to manipulate the block-diagonal Hamiltonian, for instance the diagonalisation routines:
\lstinputlisting[firstline=37, lastline=38]{\SSHcode}

Last, we proceed to calculate the correlation function from Eq.~\eqref{eq:SSH_corr}. To this end, we shall split the correlator according to the RHS of Eq.~\eqref{eq:SSH_corr}. Thus, the strategy is to evolve both the Fermi sea $|\mathrm{FS}(t)\rangle$ and the auxiliary state $|\mathrm{nFS(t)}\rangle$ in time, and compute the expectation value of the time-independent operator $n_i(0)$ in between the two states as a function of time. Keep in mind that we do not need to construct the Fermi sea as a many-body state explicitly, so we rather work with single-particle states.

The first step is to collect all momentum eigenstates into the columns of the array \texttt{psi}. We then build the operators $n_{j=0}$ and $n_{j=L/2}$ in real space.  
\lstinputlisting[firstline=39, lastline=49]{\SSHcode}
Next, we transform these two operators to momentum space using the method \texttt{rotate\_by()}. Setting the flag \texttt{generator=False} treats the Fourier transform \texttt{FT} as a unitary transformation, rather than a generator of a unitary transformation.
\lstinputlisting[firstline=50, lastline=52]{\SSHcode} 
Let us proceed with the time-evolution. We first define the time vector \texttt{t} and the state \texttt{n\_psi0}.
\lstinputlisting[firstline=53, lastline=57]{\SSHcode}
We can perform the time evolution in two ways: (i) we calculate the time-evolution operator \texttt{U} using the \texttt{exp\_op} class
\footnote{The \texttt{exp\_op} class uses the scipy functions \texttt{scipy.linalg.expm()} (when the user requests to compute the matrix exponential explicitly) and \texttt{scipy.sparse.linalg.expm\_multiply()}. Note also the existence of the QuSpin function \texttt{tools.evolution.expm\_multiply\_parallel()} which provides an OpenMP implementation of \texttt{scipy.sparse.linalg.expm\_multiply()}, but is not part of the \texttt{exp\_op} class (when only the application of the matrix exponential on a state is required).}, 
and apply it to the momentum states \texttt{psi0} and \texttt{n\_psi0}. The \texttt{exp\_op} class calculates the matrix exponential $\exp(aH)$ of an operator $H$ multiplied by a complex number $a$. One can also conveniently compute a series of matrix exponentials $\exp(aHt)$ for every time $t$ by specifying the stating point \texttt{start}, endpoint \texttt{stop} and the number of elements \texttt{num} which define the time array $t$ via \texttt{t=np.linspace(start,stop,num)}. Last, by parsing the flag \texttt{iterate=True} we create a python generator -- a pre-defined object evaluated only at the time it is called, i.e.~not pre-computed, which can save both time and memory. 
\lstinputlisting[firstline=58, lastline=62]{\SSHcode}
Another way of doing the time evolution, (ii), is to use the \texttt{evolve()} method of the \texttt{hamiltonian} class. The idea here is that every Hamiltonian defines a generator of time translations. This method solves the Schr\"odinger equation using SciPy's ODE integration routines, see App.~\ref{app:doc} for more details. The required arguments, in order of appearance, are the initial state, the initial time, and the time vector. The \texttt{evolve()} method also supports the option to create the output as a generator using the flag \texttt{iterate=True}. Both ways (i) and (ii) time-evolve all momentum states \texttt{psi} at once, i.e.~simultaneously.
\lstinputlisting[firstline=63, lastline=65]{\SSHcode}
To evaluate the correlator, we first preallocate memory by defining the empty array \texttt{correlators}, which will be filled with the correlator values in every single-particle momentum mode $|k\rangle$. Using generators allows us to loop only once over time to obtain the time-evolved states \texttt{psi\_t} and \texttt{n\_psi\_t}. In doing so, we evaluate the expectation value $\langle \mathrm{FS}(t)|n_i(0)|\mathrm{nFS}(t)\rangle$ using the \texttt{matrix\_ele()} method of the \texttt{hamiltonian} class. The flag \texttt{diagonal=True} makes sure only the diagonal matrix elements are calculated\footnote{Recall that \texttt{psi\_t} and \texttt{n\_psi\_t} contain many time-evolved states, and if one uses the default \texttt{diagonal=False}, all off-diagonal matrix elements will be computed as well, so the result will be a two-dimensional array} and returned as a one-dimensional array. 
\lstinputlisting[firstline=66, lastline=70]{\SSHcode}
Finally, we weigh all singe-state correlators by the Fermi-Dirac distribution to obtain the finite-temperature non-equal time correlation function $C_{0,L/2}(t,\beta)$.
\lstinputlisting[firstline=71, lastline=73]{\SSHcode}

The complete code including the lines that produce Fig.~\ref{fig:SSH} is available under:\\

\href{http://weinbe58.github.io/QuSpin/examples/example5.html}{http://weinbe58.github.io/QuSpin/examples/example5.html}\\

\subsection{Many-Body Localization in the Fermi-Hubbard Model}
\label{subsec:Fermion_MBL}

This example shows how to:
\begin{itemize}
	\item construct Hamiltonians for spinful fermions using the \texttt{spinful\_fermion\_basis\_1d} class,
	\item use \texttt{quantum\_operator} class to construct Hamiltonians with varying parameters,
	\item use new \texttt{basis.index()} function to construct Fock states from strings,
	\item use \texttt{obs\_vs\_time()} functionality to measure observables as a function of time.
\end{itemize}

A class of exciting new problems in the field of non-equalibrium physics is that of many-body localised (MBL) models. The MBL transition is a dynamical phase transition in the eigenstates of a many-body Hamiltonian. Driven primarily by quenched disorder, the transition is distinguished by the system having ergodic eigenstates in the weak disorder limit and non-ergodic eigenstates in the strong disorder limit. The MBL phase is reminiscent of integrable systems as one can construct quasi-local integrals of motion, but these integrals of motion are much more robust in the sense that they are not sensitive to small perturbations, as is the case in many classes of integrable systems~\cite{nandkishore_15,abanin_17,khemani_16,khemani_17,ponte_17}.

Motivated by recent experiments~\cite{kondov_15,Schreiber15,ovadia_15,bordia_16,smith_16,choi_16,kucsko_16,lueschen_17,bordia_17,zhang_17} we explore MBL in the context of fermions using QuSpin. The model we will consider is the Fermi-Hubbard model with quenched random disorder which has the following Hamiltonian:
\begin{equation}
	H = -J\sum_{i=0,\sigma}^{L-2} \left(c^\dagger_{i\sigma}c_{i+1,\sigma} - c_{i\sigma}c^\dagger_{i+1,\sigma}\right) +U\sum_{i=0}^{L-1} n_{i\uparrow }n_{i\downarrow } + \sum_{i=0,\sigma}^{L-1} V_i n_{i\sigma}\label{eq:FH_ham}
\end{equation}
where $c_{i\sigma }$ and $c^\dagger_{i\sigma }$ are the fermionic creation and annihilation operators on site $i$ for spin $\sigma$, respectively. We will work in the sector of $1/4$ filling for both up and down spins. The observable of interest is the sublattice imbalance~\cite{Schreiber15,bordia_16,lueschen_17}:
\begin{equation}
	I = (N_A-N_B)/N_\mathrm{tot}
\end{equation}
where $A$ and $B$ refer to the different sublattices of the chain and $N$ is the particle number.

We prepare an initial configuration of fermions of alternating spin on every other site. Evolving this initial state under the Hamiltonian~\eqref{eq:FH_ham}, we calculate the time dependence of the imbalance $I(t)$ which provides useful information about the ergodicity of the Hamiltonian $H$ (or the lack thereof). If the Hamiltonian is ergodic, the imbalance should decay to zero in the limit $t\rightarrow\infty$, as one would expect due to thermalising dynamics. On the other hand, if the Hamiltonian is MBL then some memory of the initial state will be retained at all times, and therefore the imbalance $I(t)$ will remain non-zero even at infinite times.

Because the Hilbert space dimension grows so quickly with the lattice size for the Fermi-Hubbard Hamiltonian, we only consider the dynamics after a finite amount of time, instead of looking at infinite-time expectation values which require knowledge of the entire basis of the Hamiltonian, which requires more computational resources. 

\begin{figure}[t!]
	\centering
	\includegraphics[width=0.5\textwidth]{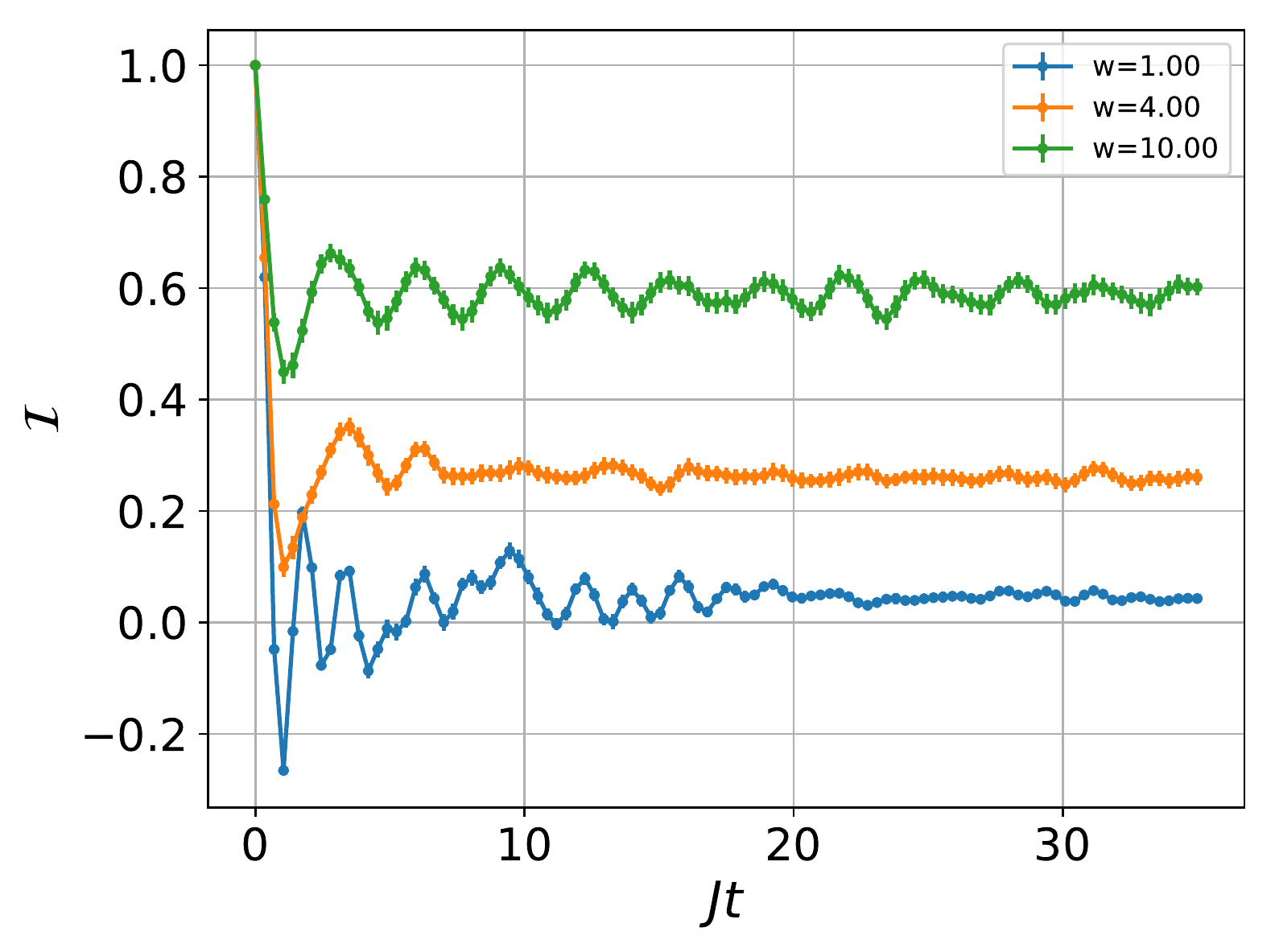}
	\caption{\label{fig:MBL} Sublattice imbalance $I$ as a function of time, averaged over $100$ disorder realizations for different disorder strengths. This data was taken on a chain of length $L=8$.}
\end{figure}

\noindent\emph{Code Analysis---}We start out by loading a set of libraries which we need to proceed with the simulation of MBL fermions. 
\lstinputlisting[firstline=2, lastline=7, firstnumber=1]{\MBLcode}
While we already encountered most of these libraries and functions, in this example we introduce the \texttt{quantum\_operator} class which defines an operator, parametrized by multiple parameters, as opposed to the Hamiltonian which is only parametrized by time. Also, since this example requires us to do many different disorder realisations, we use \texttt{NumPy}'s random number library to do random sampling. We load \texttt{uniform} to generate the uniformly distributed random potential as well as \texttt{choice} which we use to estimate the uncertainties of the disorder averages using a bootstrap re-sampling procedure which we explain below. In order to time how long each realization takes, we use the \texttt{time} function from python's \texttt{time} library. 

After importing all the required libraries and functions we set up the parameters for the simulation including the number of realizations \texttt{n\_real}, and the physical couplings \texttt{J}, \text{U}, the number of up and down fermions, etc.
\lstinputlisting[firstline=10, lastline=24, firstnumber=10]{\MBLcode}
Next we set up the basis, introducing here the \texttt{spinful\_fermion\_basis\_1d} constructor class. It works the usual way, except the particle number sector \texttt{N\_f} is now a tuple containing the number of up and down fermions. 
\lstinputlisting[firstline=26, lastline=28,firstnumber=26]{\MBLcode}
The next step in the procedure is to set up the site-coupling and operator lists:
\lstinputlisting[firstline=30, lastline=45,firstnumber=30]{\MBLcode}
Notice here that the \texttt{"|"} character is used to separate the operators which belong to the up (left side of tensor product) and down (right side of tensor product) Hilbert spaces in \texttt{basis}. If no string in present the operator is assumed to be the identity \texttt{'I'}. The site-coupling list, on the other hand, does not require separating the two sides of the tensor product, as it is assumed that the operator string lines up with the correct site index when the \texttt{'|'} character is removed. 

In the last couple of lines defining model (see below), we create a python \texttt{dictionary} object in which we add static operator lists as \texttt{values}, indexed by a particular string known as the corresponding \texttt{key}. This dictionary, which we refer to as \texttt{operator\_dict}, is then passed into the \texttt{quantum\_operator} class which, for each \texttt{key} in \texttt{operator\_dict}, constructs the operator listed in the site-coupling list for that \texttt{key}. When one wants to evaluate this operator for a particular set of parameters, one uses a second dictionary (see \texttt{params\_dict} below in lines $74-76$) with the same keys as \texttt{operator\_dict}: the value corresponding to each key in the \texttt{params\_dict} multiples the operator stored at that same key in \texttt{operator\_dict}. This allows one to parametrize many-body operators in more complicated and general ways. In the present example, we define a key for the Fermi-Hubbard Hamiltonian, and then, as we need to change the disorder strength in between realisations, we also define keys for the local density operator for both up and down spins on each site. By doing this, we can then construct any disordered Hamiltonian by specify the disorder at each site, and changing its value from one realisation to another.
\lstinputlisting[firstline=46, lastline=52,firstnumber=46]{\MBLcode}

The \texttt{quantum\_operator} class constructs operators in almost an identical manner as a \texttt{hamiltonian()} class with the exception that there is no dynamic list. Next, we construct our initial state with the fermions dispersed over the lattice on every other site. To get the index of the \texttt{basis} state which this initial state corresponds to, one can use the \texttt{index} function of the \texttt{tensor\_basis} class. This function takes a string or integer representing the product state for each of the Hilbert spaces and then searches to find the full product state, returning the corresponding index. We can then create an empty array \texttt{psi\_0} of dimension the total Hilbert space size, and insert unity at the index corresponding to the initial state. This allows us to easily define the many-body product state used in the rest of the simulation.
\lstinputlisting[firstline=54, lastline=67,firstnumber=54]{\MBLcode}

Now that the operators are all set up, we can proceed with the simulation of the dynamics. First, we define a function which, given a disorder realization of the local potential, calculates the time evolution of $\mathcal{I}(t)$. We shall guide the reader through this function step by step. The syntax for this begins as follows:
\lstinputlisting[firstline=69, lastline=72,firstnumber=69]{\MBLcode}
The first step is to construct the Hamiltonian from the \texttt{disorder} list which is as simple as
\lstinputlisting[firstline=73, lastline=79,firstnumber=73]{\MBLcode}
using the \texttt{tohamiltonian()} method of \texttt{H\_dict} class, which accepts as argument the dictionary \texttt{params\_dict}, which shares the same keys as \texttt{operator\_dict}, but whose values are determined by the \texttt{disorder} list which changes from one realisation to another. 

Once the Hamiltonian has been constructed we want to time-evolve the initial state with it. To this end, we use the fact that, for time-independent Hamiltonians, the time-evolution operator coincides with the matrix exponential $\exp(-itH)$. In QuSpin, a convenient way to define matrix exponentials is offered by the \texttt{exp\_op} class. Given an operator $A$ , it calculates $\exp(atA)$ for any complex-valued number $a$. The time grid for $t$ is specified using the optional arguments \texttt{start}, \texttt{stop} and \texttt{num}. If the latter are omitted, default is $t=1$. The \texttt{exp\_op} objects are either a list of arrays, the elements of which correspond to the operator $\exp(atA)$ at the times defined or, if \texttt{iterate=True} a generator for this list is returned. Here, we use \texttt{exp\_op} to create a generator list containing the time-evolved states as follows
\lstinputlisting[firstline=80, lastline=82,firstnumber=80]{\MBLcode}

To calculate the expectation value of the imbalance operator in time, we use the \texttt{obs\_vs\_time} function. Since the usage of this function was extensively discussed in \href{http://weinbe58.github.io/QuSpin/examples/example2.html}{Example 2} and \href{http://weinbe58.github.io/QuSpin/examples/example3.html}{Example 3} of Ref.~\cite{weinberg_17_quspin}, here we only mention that it accepts a (collection of) state(s) [or a generator] \texttt{psi\_t}, a time vector \texttt{t}, and a dictionary \texttt{dict(I=I)}, whose values are the observables to calculate the expectation value of. The function returns a dictionary, the keys of which correspond to the keys every observable was defined under.
\lstinputlisting[firstline=83, lastline=85,firstnumber=83]{\MBLcode}

The function ends by printing the time of executing and returning the value for $I$ as a function of time for this realization
\lstinputlisting[firstline=86, lastline=89,firstnumber=86]{\MBLcode}

Now we are all set to run the disorder realizations for the different disorder strengths which in principle can be spilt up over multiple simulations, e.g.~using \texttt{joblib} [c.f.~\href{http://weinbe58.github.io/QuSpin/examples/example1.html}{Example 1} from Ref.~\cite{weinberg_17_quspin}] but for completeness we do all of the calculations in one script. 
\lstinputlisting[firstline=91, lastline=93,firstnumber=91]{\MBLcode}
Last, we calculate the average and its error bars using \emph{bootstrap re-sampling}. The idea is that we have a given set of \texttt{n\_real} samples from which we can select randomly a set of \texttt{n\_real} samples with replacements. This means that sometimes we will get the same individual sample represented multiple times in one of these sets. Then by averaging over this set one can get an estimate of the mean value for the original sample set. This mean value is what is called a bootstrap sample. In principle there are a very large number of possible bootstrap samples so in practice one only takes a small fraction of the total set of bootstrap samples from which you can estimate things like the variance of the mean which is the error of the original sample set.
\lstinputlisting[firstline=94, lastline=100,firstnumber=94]{\MBLcode}

The complete code including the lines that produce Fig.~\ref{fig:MBL} is available under:\\

\href{http://weinbe58.github.io/QuSpin/examples/example6.html}{http://weinbe58.github.io/QuSpin/examples/example6.html}\\

\subsection{The Bose-Hubbard Model on Translationally Invariant Ladder}
\label{subsec:Bose_Ladder}

This example shows how to:
\begin{itemize}
	\item construct interacting Hamiltonians for bosonic systems,
	\item construct a Hamiltonian on a ladder geometry,
	\item use the \texttt{block\_ops} class to evolve a state over several symmetry sectors at once,
	\item measure the entanglement entropy of a state on the ladder.
\end{itemize}

\noindent{\emph{Physics Setup---}} In this example we use QuSpin to simulate the dynamics of the Bose-Hubbard model (BHM) on a ladder geometry. The BHM is a minimal model for interacting lattice bosons~\cite{sachdev_book} which is most often experimentally realizable in cold atom experiments~\cite{bloch_05}. The Hamiltonian is given by
\begin{equation}
	H_\mathrm{BHM} = -J\sum_{\langle ij\rangle} \left(b_i^\dagger b_j + \mathrm{h.c.}\right) + \frac{U}{2}\sum_{i}n_i(n_i-1)
	\label{eq:BHM_ham}
\end{equation} 
where $b_i$ and $b^\dagger_i$ are bosonic creation and annihilation operators on site $i$, respectively, and $\langle ij\rangle$ denotes nearest neighbors on the ladder. We consider a half-filled ladder of length $L$ with $N=2L$ sites and cylindrical boundary condition, i.e.~a periodic boundary condition along the ladder-leg direction. We are interested in the strongly-interacting limit $U/J\gg 1$, where the mean-field Gross-Pitaevskii theory fails. Hence, we restrict the local Hilbert space to allow at most two particles per site, effectively using a total of three states per site: empty, singly and doubly occupied.  

The system is initialised in a random Fock state, and then evolved with the Hamiltonian~\eqref{eq:BHM_ham}.  Since the BHM is non-integrable, we expect that the system eventually thermalizes, so that the long-time occupation becomes roughly uniform over the entire system. Besides the local density, we also measure the entanglement entropy between the two legs of the ladder.

\begin{figure}[t!]
	\centering
	\begin{subfigure}[b]{0.496\textwidth}
		\includegraphics[width=\textwidth]{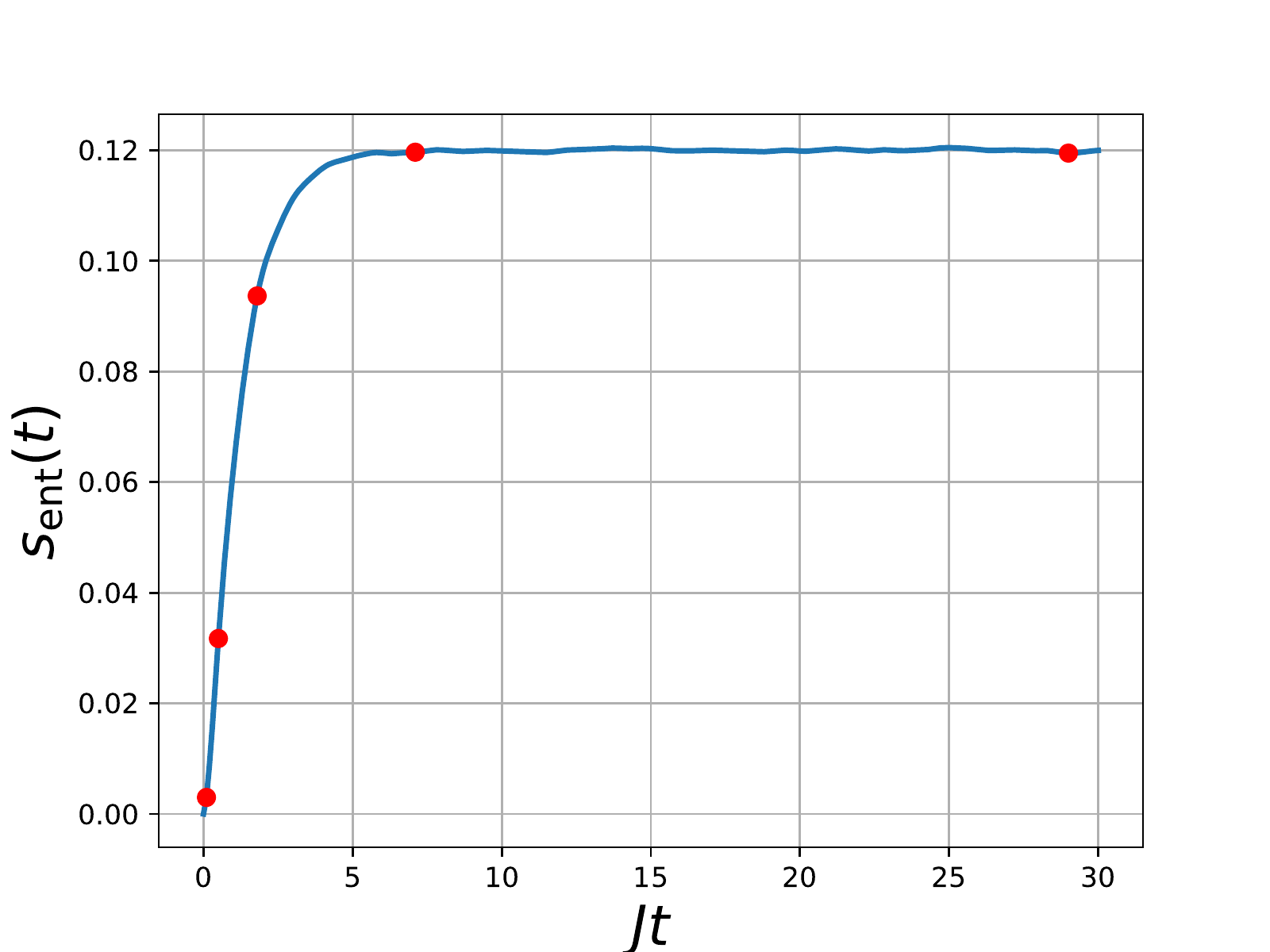}
		\caption{Entanglement entropy density between the two legs of the ladder as a function of time.}
	\end{subfigure}
	\begin{subfigure}[b]{0.496\textwidth}
		\includegraphics[width=\textwidth]{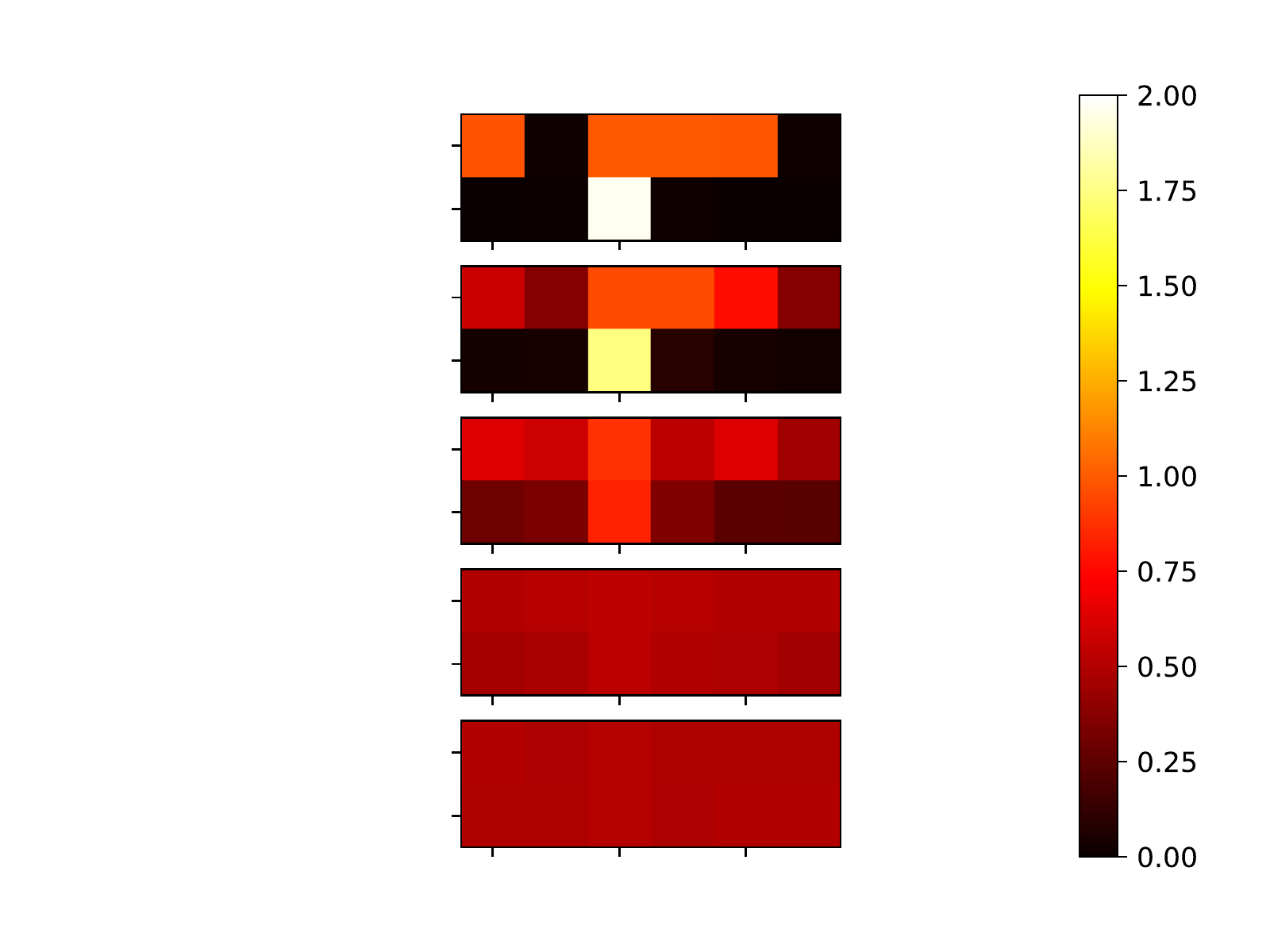}
		\caption{Local density of bosons as a function of time. Later times appearing farther down the page.}
	\end{subfigure}
	\caption{\label{fig:BHM} Quantities measured in the BHM: (a) the leg-to-leg entanglement entropy density and (b) the local density on each site as a function of time. The red dots in (a) show the time points at which the snapshots of the local density (b) are taken. The parameters are $U/J=20$, and $L=6$.}
\end{figure}

As we consider a translational invariant ladder, the Hilbert space factorizes into subspaces corresponding to the different many-body momentum blocks. This is similar to what was discussed in Sec.~\ref{subsec:SSH_model} for the SSH model, but slightly different as we consider translations of the \emph{many-body} Fock states as opposed to the single particle states~\cite{anders10}. In certain cases, it happens that, even though the Hamiltonian features symmetries, the initial state does not obey them, as is the case in the present example, where the initial state is a random Fock state. Thus, in this section we project the wavefunction to the different symmetry sectors and evolve each of the projections separately under the Hamiltonian for that symmetry sector. After the evolution, each of these symmetry-block wavefunctions is transformed back to the local Fock space basis, and summed up to recover the properly evolved initial state. We can then measure quantities such as the on-site density and the entanglement entropy. Figure~\ref{fig:BHM}(a) show the entanglement density between the two legs as a function of time, while Figure~\ref{fig:BHM}(b) shows a heat map of the local density of the bosonic gas. Both quantities show that after a short period of time the gas is completely thermalized. 

\noindent\emph{Code Analysis---}Let us now show how one can simulate this system using QuSpin. Following the code structure of previous examples, we begin by loading the modules needed for the computation:
\lstinputlisting[firstline=1, lastline=7, firstnumber=1]{\BHLcode}
First, we set up the model parameters defining the length of the ladder \texttt{L}, the total number of sites in the chain \texttt{N=2*L}, as well as the filling factor for the bosons \texttt{nb}, and the maximum number of states per site (i.e.~the local on-site Hilbert space dimension) \texttt{sps}. The hopping matrix elements $J_\perp$, $J_{\parallel,1}$, and $J_{\parallel,2}$ (see Code Snippet~\ref{snippet:BHL_geometry}) correspond to the python script variables \texttt{J\_perp}, \texttt{J\_par\_1}, and \texttt{J\_par\_2}, respectively. The on-site interaction is denoted by \texttt{U}.
\lstinputlisting[firstline=9, lastline=20, firstnumber=9]{\BHLcode}
Let us proceed to construct the Hamiltonian and the observables for the problem. For bosonic systems, we have \texttt{'+'},\texttt{'-'} \texttt{'n'}, and \texttt{'I'} as available Fock space operators to use. In order to set up the local Hubbard interaction, we first split it up in two terms: $U/2\sum_i n_i(n_i-1) = -U/2\sum_i n_i + U/2\sum_i n_i^2$. Thus, we need two coupling lists:
\lstinputlisting[firstline=22, lastline=25, firstnumber=22]{\BHLcode}

\lstinputlisting[label=snippet:BHL_geometry,caption=Translationally Invariant Ladder Graph and its Symemtries,firstline=117,lastline=141,numbers=none]{\BHLcode}

We also define the hopping site-coupling list. In general, QuSpin can set up the Hamiltonian on any graph (thus, including higher-dimensions). By labelling the lattice sites conveniently, we can define the ladder geometry, as shown in Code Snippet~\ref{snippet:BHL_geometry}. For the current ladder geometry of interest, there are three types of hopping: one along each of the two legs with tunnelling matrix elements \texttt{J\_par\_1} and \texttt{J\_par\_1}, respectively, and the transverse hopping along the rungs of the ladder with strength \texttt{J\_perp}. The even sites correspond to the bottom leg while the odds sites are the top leg of the ladder. Therefore, the hopping on the bottom/top leg is defined by \texttt{[J\_par\_...,i,(i+2)\%N]}, while the rung hopping (from the top leg to the bottom leg) is defined by \texttt{[J\_perp,i,(i+1)\%N]}.
\lstinputlisting[firstline=26, lastline=30, firstnumber=26]{\BHLcode}
where we used the \texttt{list\_1.extend(list\_2)} method to concatenate two lists together\footnote{Note that the \texttt{extend} function is done inplace so if one tries to do \texttt{new\_list=list\_1.extend(list\_2)}, \texttt{new\_list} will be \texttt{None} and \texttt{list\_1} will have all of the elements of \texttt{list\_2} appended to it.}.
 
Next, we define the static and dynamic lists, which are needed to construct the Hamiltonian. 
\lstinputlisting[firstline=31, lastline=38, firstnumber=31]{\BHLcode}
Instead of creating a \texttt{hamiltonian} class object in real space, we use the \texttt{block\_ops} class to set up the Hamiltonian in a block-diagonal form in momentum space, similar to the SSH model, c.f.~Sec.~\ref{subsec:SSH_model}, but now genuinely many-body. In order to reduce the computational cost, the state is evolved in momentum space and projected to the Fock basis after the calculation. The purpose of \texttt{block\_ops} is to provide a simple interface for solving the Schr\"odinger equation when an initial state does not obey the symmetries of the Hamiltonian it is evolved under. We have seen an example of this in Sec.~\ref{subsec:SSH_model} when trying to measure non-equal space-time correlation functions of local operators in a translational invariant system, while in this section we explicitly start out with a state which does not obey translation invariance. To construct the \texttt{block\_ops} object we use the follow set of lines, explained below:
\lstinputlisting[firstline=39, lastline=45, firstnumber=39]{\BHLcode}
First, we create a list of dictionaries \texttt{blocks} which defines the different symmetry sectors to project the initial state to, before doing the time evolution\footnote{\texttt{block\_ops} will not evolve in those symmetry sectors for which the projection is $0$.}. The optional arguments \texttt{basis\_args} and \texttt{basis\_kwargs} apply to every symmetry sector. Last, \texttt{get\_proj\_kwargs} contains the optional arguments to construct the projectors\footnote{In this case setting \texttt{pcon=True} means that the projector takes the state \emph{from} the symmetry reduced basis \emph{to} the fixed particle number basis.}. This class automatically projects the initial state onto the different symmetry sectors and will evolve the sectors individually in serial or with an additional option in parallel over multiple cpu cores. After the evolution the class reconstructs the state and returns it back to the user. The Hamiltonian and the evolution itself will only be calculated for the sectors which have non-vanishing overlap with the initial state. Furthermore, the class will, by default, construct things on the fly as it needs them to save the user memory. For more information about this class we refer the user to the Documentation, c.f.~App.~\ref{app:doc}.

Finally, we define the local density operators on the full particle-conserving Hilbert space using the \texttt{boson\_basis\_1d} class.
\lstinputlisting[firstline=46, lastline=50, firstnumber=46]{\BHLcode}

Having set up the Hamiltonian, we now proceed to the time-evolution part of the problem. We begin by defining the initial random state in the Fock basis.
\lstinputlisting[firstline=52, lastline=58, firstnumber=52]{\BHLcode}
Next we define the times which we would like to solve the Schrödinger equation for. Since the Hamiltonian is time-independent, we use the \texttt{exp\_op} class to compute the time-evolution operator as the exponential of the Hamiltonian, see Sec.~\ref{subsec:Fermion_MBL}. Therefore, we consider linearly spaced time points defined by the variables \texttt{start},\texttt{stop}, and \texttt{num}. 
\lstinputlisting[firstline=59, lastline=61, firstnumber=59]{\BHLcode}
To calculate the states as a function of time, we use the \texttt{expm} function of the \texttt{block\_ops} class to first construct the unitary evolution operator as the matrix exponential of the time-independent Hamiltonian\footnote{For time dependent Hamiltonians, the \texttt{block\_ops} class contains a method called \texttt{evolve}, see App.~\ref{app:doc}.}. We define the \texttt{expm} function to have almost identical arguments as that of the \texttt{exp\_op} class, but with some major exceptions. For one, because the Hamiltonian factorizes in a block-diagonal form, the evolution over each block can be done separately (e.g.~just trivially loop through the blocks sequentially). In some cases, however, e.g.~for single particle Hamiltonians [see Sec.~\ref{subsec:SSH_model}], there are a lot of small blocks, and it actually makes sense to calculate the matrix exponential in block diagonal form which is achieved by setting the optional argument \texttt{block\_diag=True}. Another optional argument, \texttt{n\_jobs=int}, allows the user to spawn multiple python processes which do the calculations for the different blocks simultaneously for \texttt{n\_jobs>1}. On most systems these processes will be distributed over multiple CPUs which can speed up the calculations if there are resources for this available. This also works in conjunction with the \texttt{block\_diag} flag where each process creates its own block diagonal matrix for the calculation. Once all the calculations for each block are completed, the results are combined and conveniently projected back to the original local Fock basis automatically. 
\lstinputlisting[firstline=62, lastline=64, firstnumber=62]{\BHLcode}
We can now use the time dependent states calculated to compute the expectation value of the local density
\lstinputlisting[firstline=65, lastline=70, firstnumber=65]{\BHLcode}
We can also compute the entanglement entropy between the two legs of the ladder. In the newer versions of QuSpin we have moved the entanglement entropy calculations to the \texttt{basis} classes themselves (keeping of course backwards compatible functions from older versions). Once again, we refer the reader to the Documentation to learn more about how to use this function, see App.~\ref{app:doc}. 
\lstinputlisting[firstline=71, lastline=74, firstnumber=71]{\BHLcode}

The complete code including the lines that produce Fig.~\ref{fig:BHM} is available under:\\

\href{http://weinbe58.github.io/QuSpin/examples/example7.html}{http://weinbe58.github.io/QuSpin/examples/example7.html}\\

\subsection{The Gross-Pitaevskii Equation and Nonlinear Time Evolution}
\label{subsec:GP_dynamics}

This example shows how to:
\begin{itemize}
	\item simulate user-defined time-dependent nonlinear equations of motion using the \\ \texttt{evolution.evolve()} routine,
	\item use imaginary time dynamics to find the lowest energy state.
\end{itemize}

\noindent\emph{Physics Setup---}The Gross-Pitaevskii wave equation (GPE) has been shown to govern the physics of weakly-interacting bosonic systems. It constitutes the starting point for studying Bose-Einstein condensates~\cite{dalfovo_99}, but can also appear in non-linear optics\cite{sulem2007nonlinear}, and represents the natural description of Hamiltonian mechanics in the wave picture~\cite{polkovnikov2010phase}. One of its interesting features is that it can exhibit chaotic classical dynamics, a physical manifestation of the presence of a cubic non-linear term.

Here, we study the time-dependent GPE on a one-dimensional lattice:
\begin{eqnarray}
i\partial_t\psi_j(t) &=& -J\left( \psi_{j-1}(t) + \psi_{j+1}(t)\right) + \frac{1}{2}\kappa_\mathrm{trap}(t)(j-j_0)^2\psi_j(t) + U|\psi_j(t)|^2\psi_j(t), \nonumber \\
\kappa_\mathrm{trap}(t)&=&(\kappa_f-\kappa_i)t/t_\mathrm{ramp}+ \kappa_i
\label{eq:GPE}
\end{eqnarray}
where $J$ is the hopping matrix element, $\kappa_\mathrm{trap}(t)$ -- the harmonic trap strength which varies slowly in time over a scale $t_\mathrm{ramp}$, and $U$ -- the mean-field interaction strength. The lattice sites are labelled by $j=0,\dots,L-1$, and $j_0$ is the centre of the $1d$ chain. We set the lattice constant to unity, and use open boundary conditions.

Whenever $U=0$, the system is non-interacting and the GPE reduces to the Heisenberg EOM for the bosonic field operator $\hat\psi_j(t)$. Thus, for the purposes of using QuSpin to simulate the GPE, it is instructive to cast Eq.~\eqref{eq:GPE} in the following generic form
\begin{equation}
i\partial_t\vec{\psi}(t) = H_\mathrm{sp}(t)\vec{\psi}(t) + U \vec{\psi}^*(t)\circ \vec{\psi}(t)\circ \vec{\psi}(t), 
\end{equation}  
where $[\vec{\psi}(t)]_j = \psi_j(t)$, and $\circ$ represents the element-wise multiplication 
\begin{equation*}
\vec{\psi}(t)\circ \vec{\phi}(t) = \bigg(\psi_0(t)\phi_0(t), \psi_1(t)\phi_1(t),\dots, \psi_{L-1}(t)\phi_{L-1}(t) \bigg)^t.
\end{equation*}
The time-dependent single-particle Hamiltonian in real space is represented as an $L\times L$ matrix, $H_\mathrm{sp}(t)$, which comprises the hopping term, and the harmonic trap.

\begin{figure}[t!]
	\centering
	
	\begin{subfigure}[b]{0.325\textwidth}
		\includegraphics[width=\textwidth]{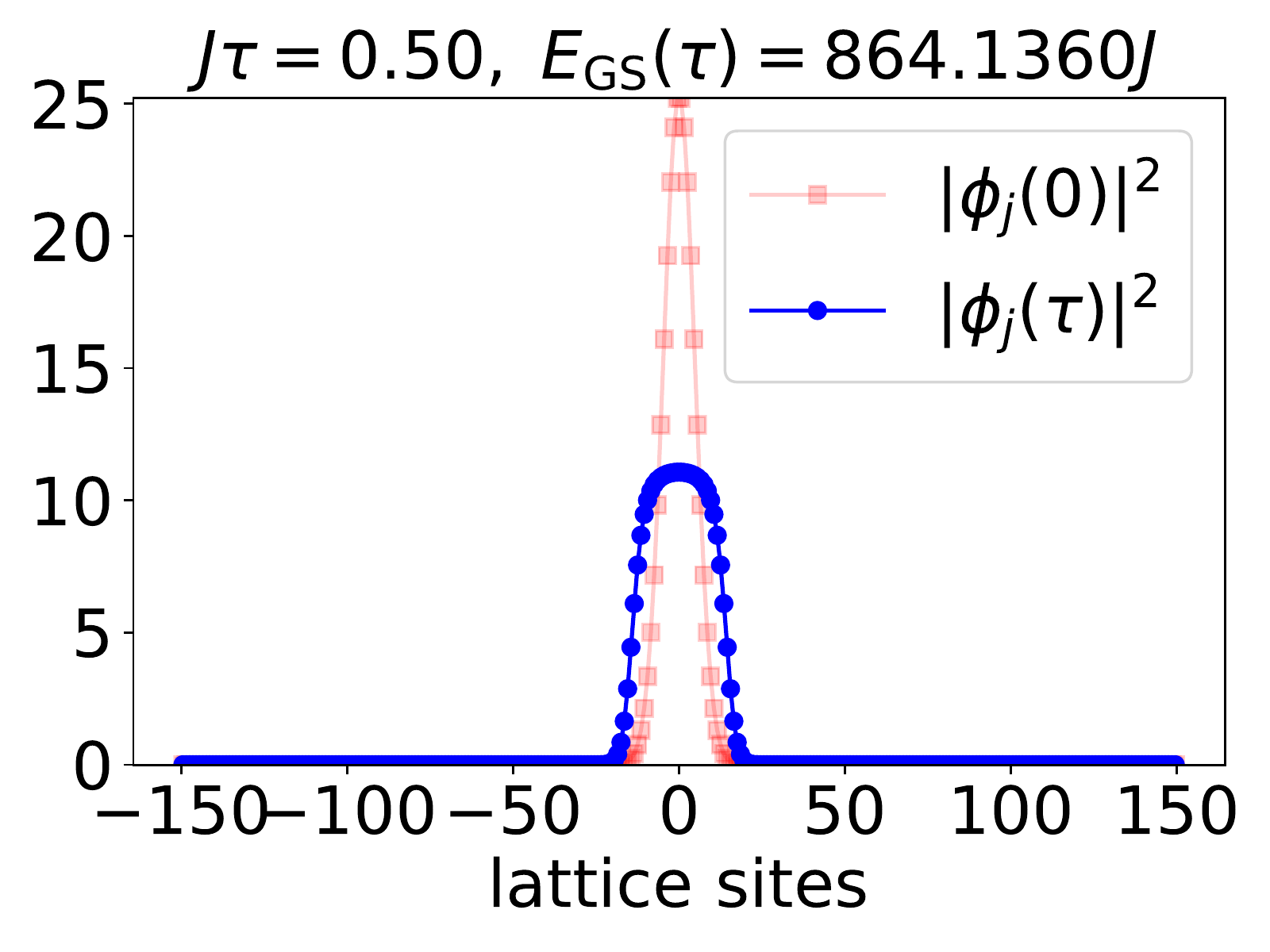}
	\end{subfigure}
	\begin{subfigure}[b]{0.325\textwidth}
		\includegraphics[width=\textwidth]{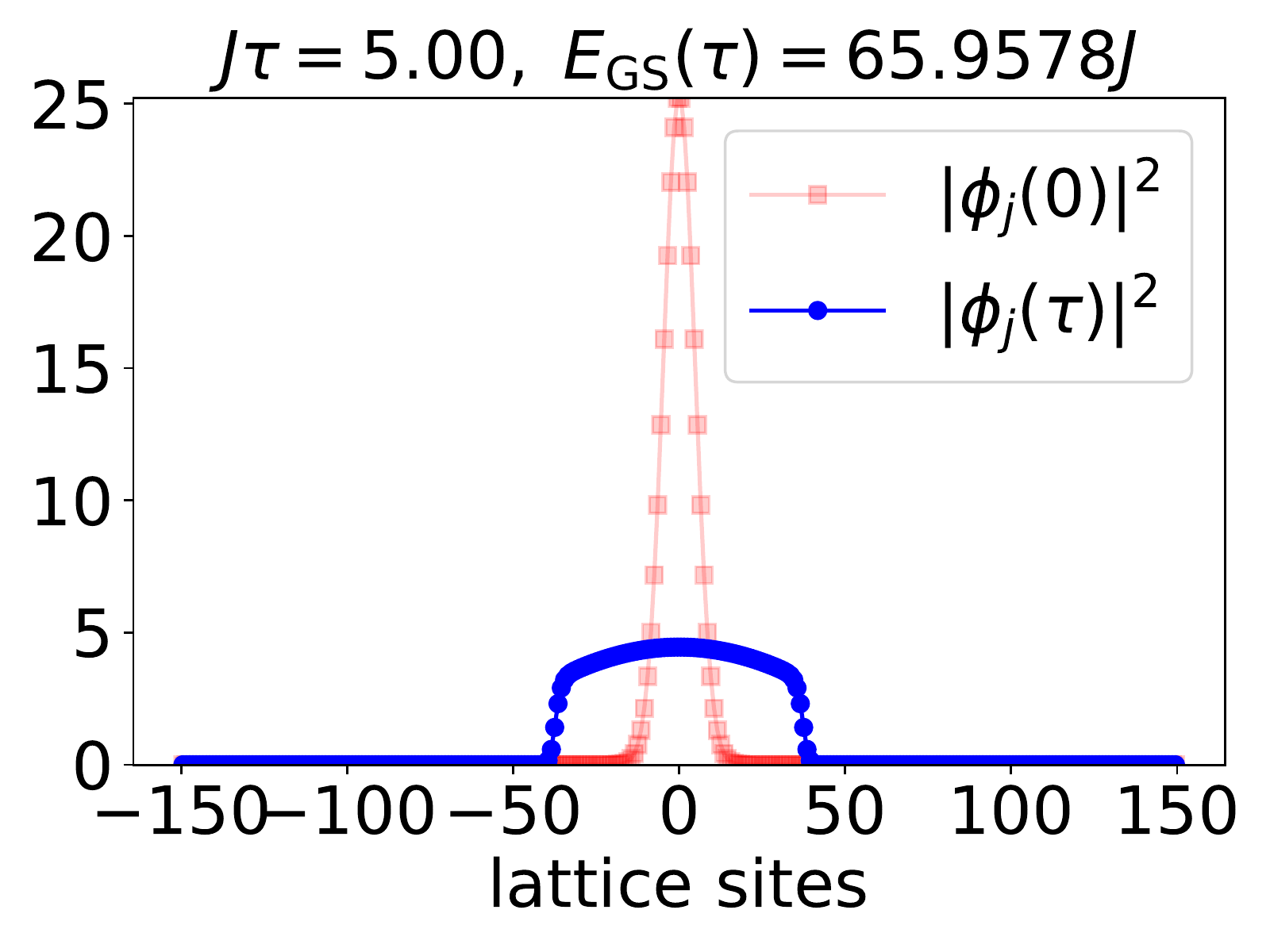}
	\end{subfigure}
	\begin{subfigure}[b]{0.325\textwidth}
		\includegraphics[width=\textwidth]{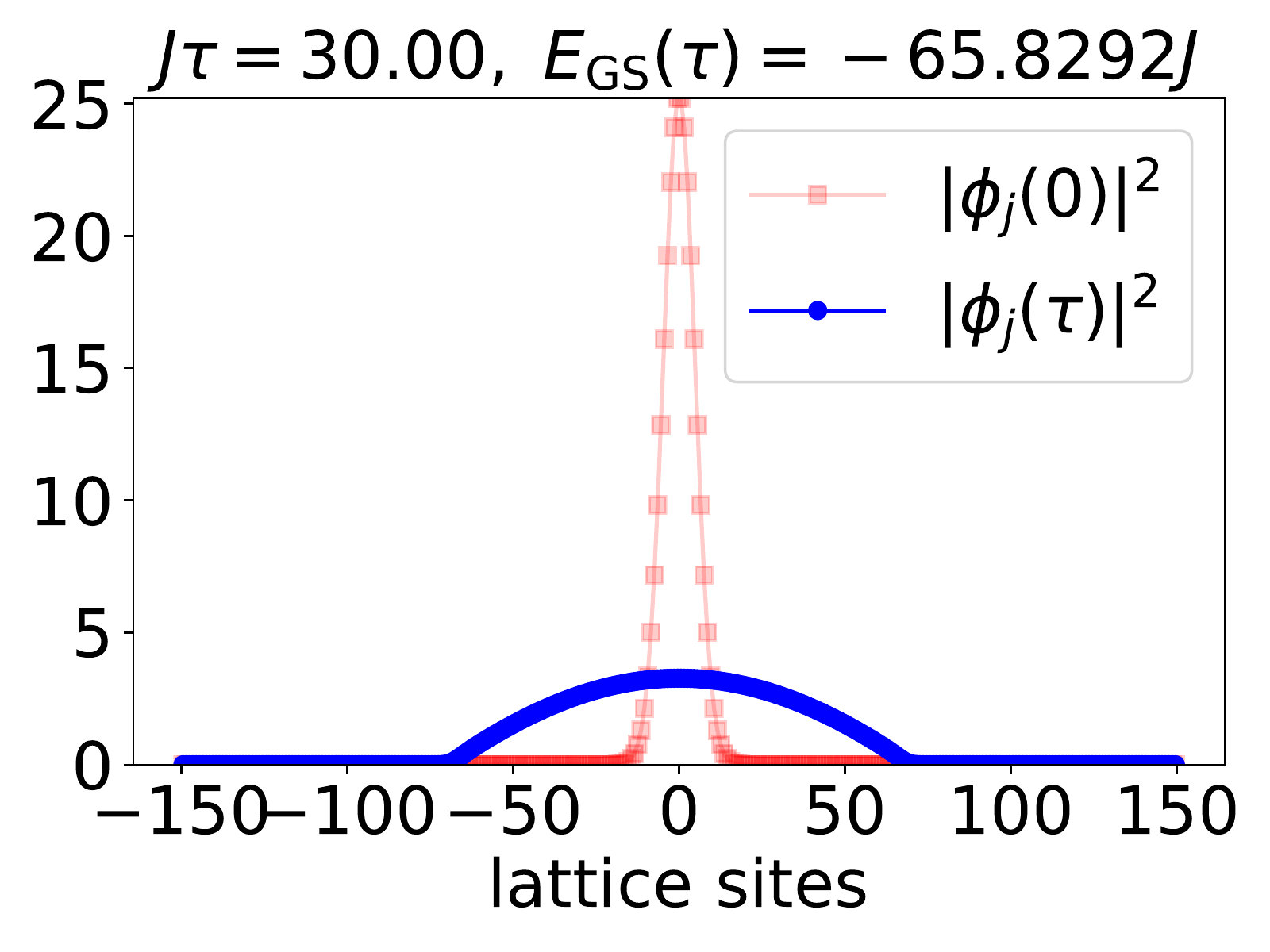}
	\end{subfigure}
	
	\begin{subfigure}[b]{0.325\textwidth}
		\includegraphics[width=\textwidth]{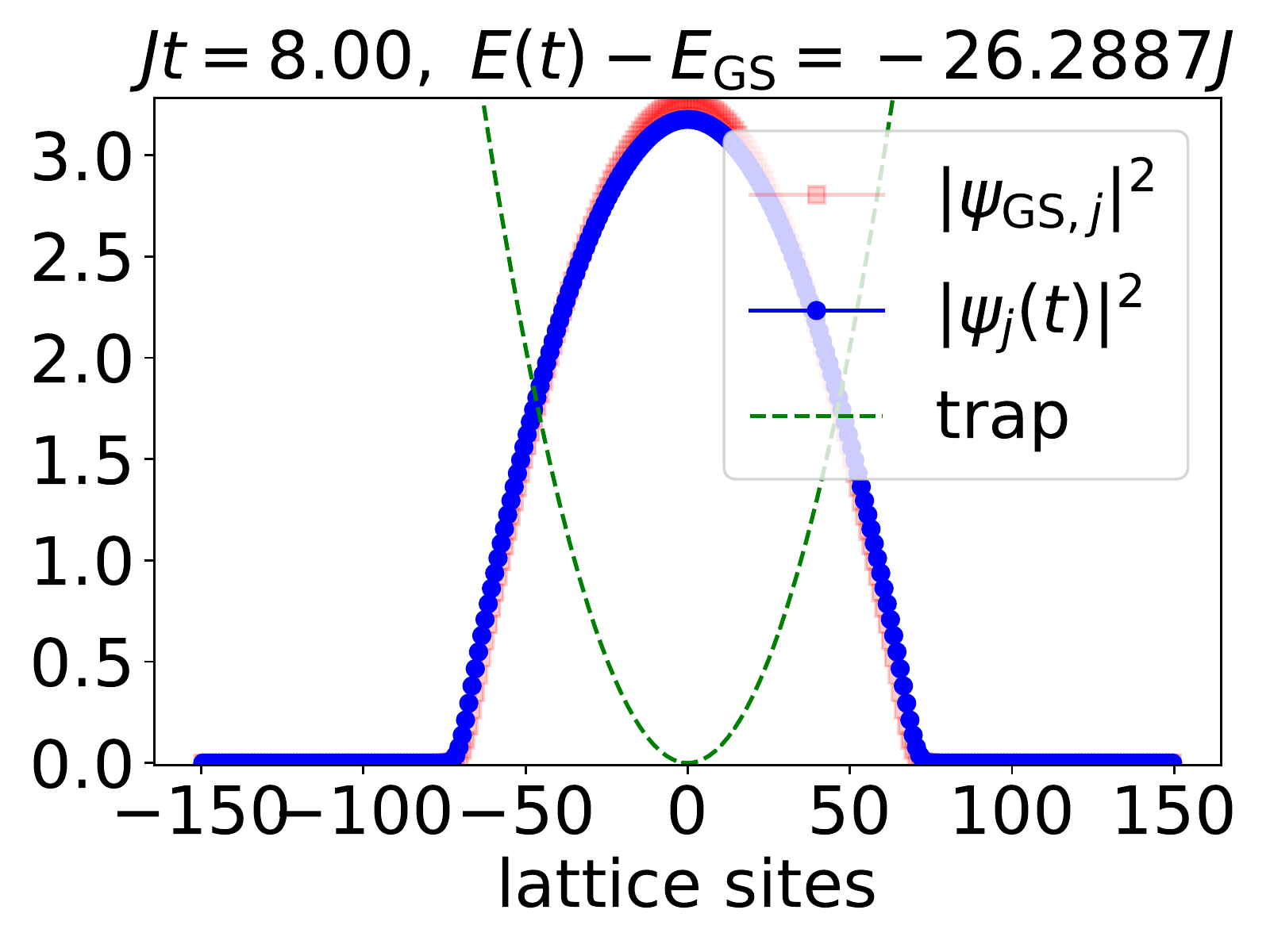}
	\end{subfigure}
	\begin{subfigure}[b]{0.325\textwidth}
		\includegraphics[width=\textwidth]{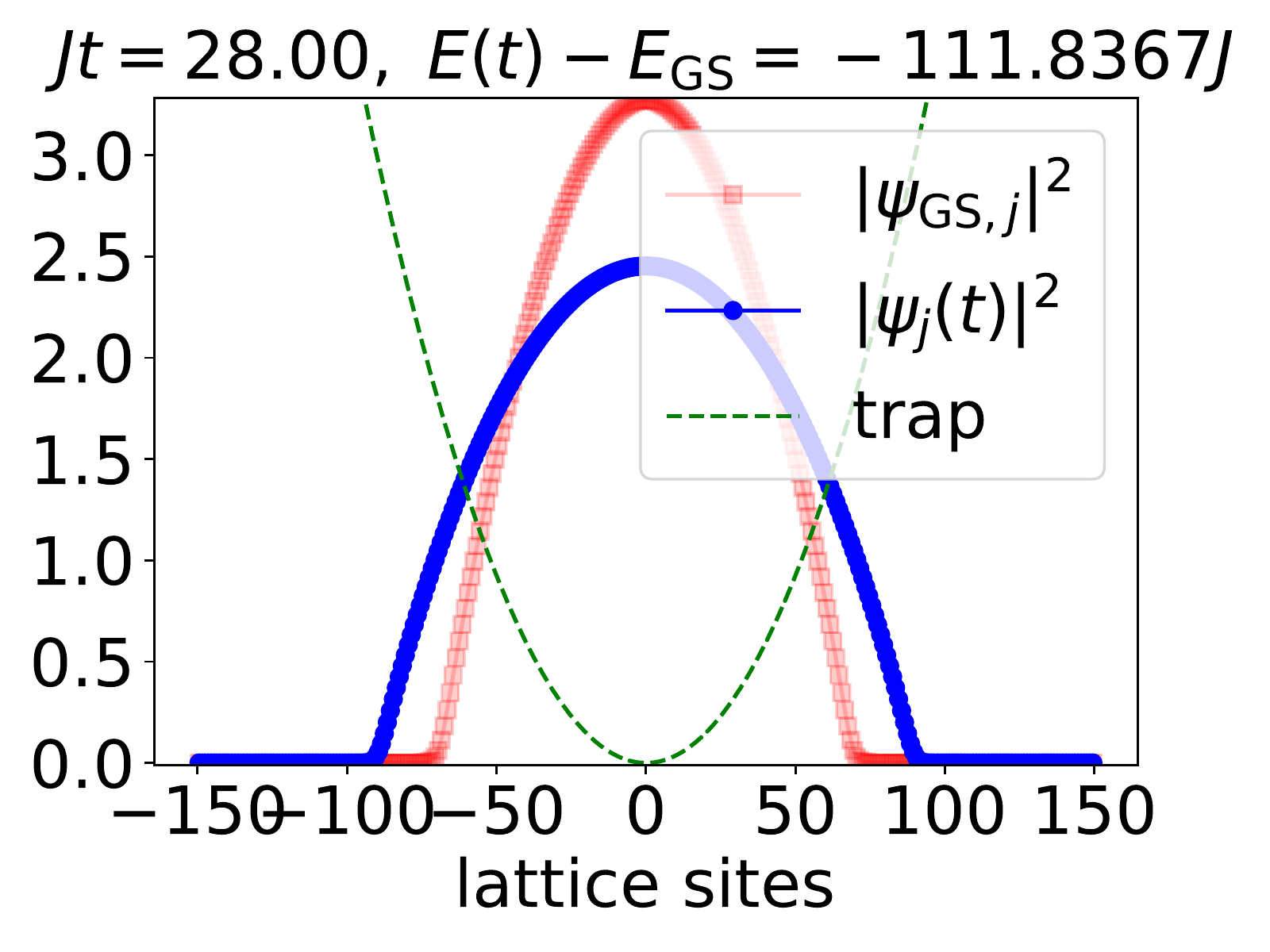}
	\end{subfigure}
	\begin{subfigure}[b]{0.325\textwidth}
		\includegraphics[width=\textwidth]{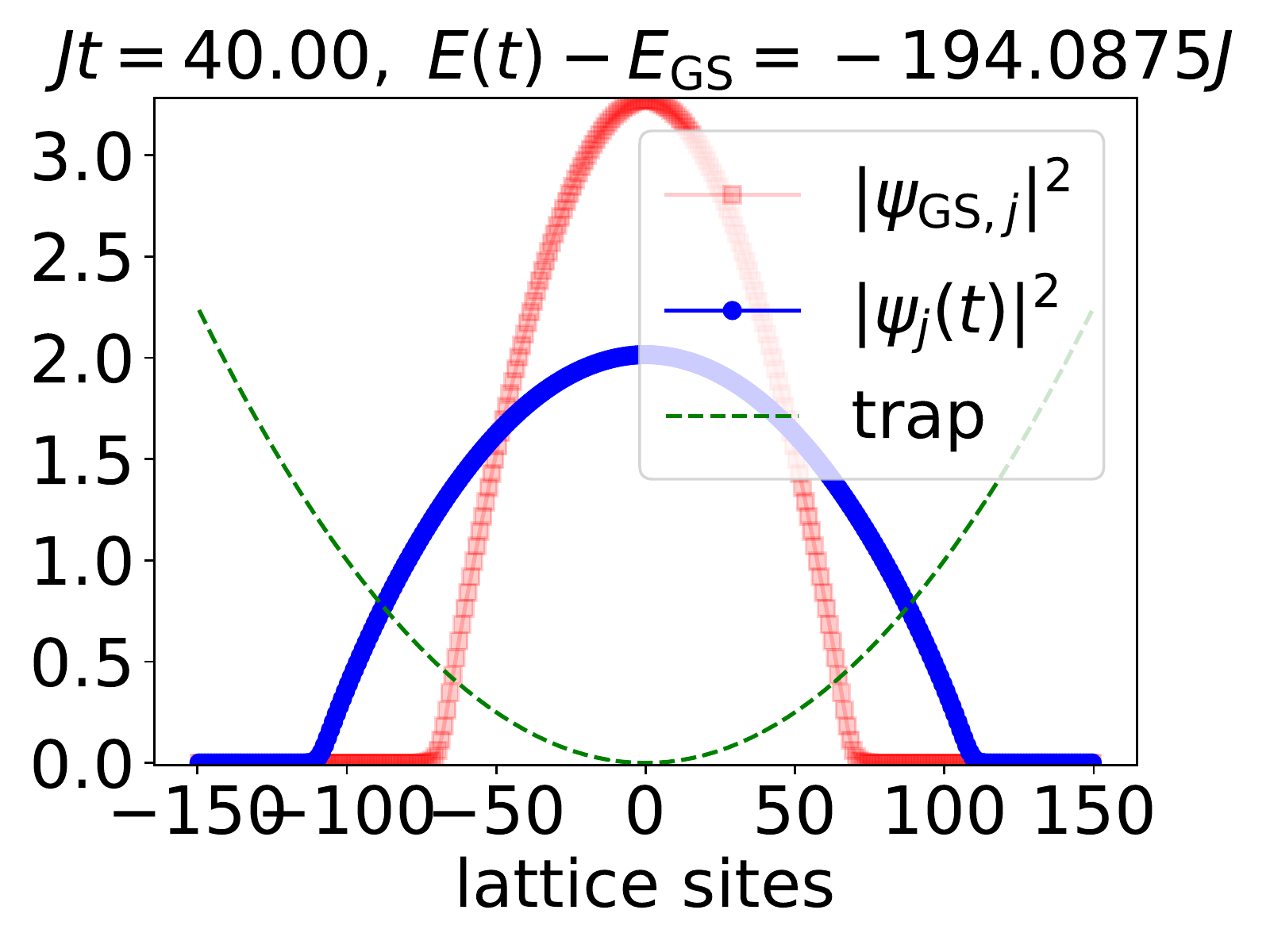}
	\end{subfigure}
	
	\caption{\label{fig:GPE_fig} Top row: three snapshots of the imaginary time dynamics which turns the non-interacting ground state into the GPE ground state at long imaginary times. Bottom row: three snapshots of the real-time evolution induced by slowly widening the harmonic trap strength. The pale red curve is the initial state, while the blue curve shows the instantaneous state. The harmonic trap is shown with a dashed green line.}  
\end{figure}

We want to initiate the time-evolution of the system at $t=0$ in its lowest energy state. To this end, we can define a `ground state' for the GPE equation, in terms of the configuration which minimises the energy of the system:
\begin{eqnarray}
\vec\psi_\mathrm{GS} &=& \inf_{\vec{\psi}} \bigg( \vec\psi^t H_\mathrm{sp}(0)\vec{\psi} + \frac{U}{2}\sum_{j=0}^{L-1}|\psi_j|^4\bigg),\nonumber\\
&=&\inf_{\psi_j} \bigg(-J\sum_{j=0}^{L-2} (\psi_{j+1}^*\psi_j + \mathrm{c.c.}) + \frac{1}{2}\kappa_\mathrm{trap}(0)\sum_{j=0}^{L-1}(j-j_0)^2|\psi_j|^2 + \frac{U}{2}\sum_{j=0}^{L-1}|\psi_j|^4\bigg).
\end{eqnarray} 
One way to find the configuration $\vec\psi_\mathrm{GS}$, is to solve the GPE in imaginary time ($it\to \tau$), which induces exponential decay in all modes of the system, and singles out the lowest-energy state in the longer run. In doing so, we keep the norm of the solution fixed:
\begin{eqnarray}
\label{eq:GPE_imag}
\partial_{\tau}\vec\varphi(\tau) &=& -\bigg[H_\mathrm{sp}(0)\vec\varphi(\tau) + U {\vec\varphi}^*(\tau)\circ \vec\varphi(\tau)\circ \vec\varphi(\tau)\bigg],\qquad ||\vec\varphi(\tau)||=\mathrm{const.},\nonumber\\
\vec{\psi}_\mathrm{GS} &=& \lim_{\tau\to\infty}\vec\varphi(\tau)
\end{eqnarray}
Snapshots of the imaginary-time evolution can be seen in Fig.~\eqref{fig:GPE_fig}, top row.

Once we have the initial state $\vec\psi_\mathrm{GS}$, we evolve it according to the time-dependent GPE, Eq.~\eqref{eq:GPE}, and track down the time evolution of the condensate density $\rho_j(t)=|\psi_j(t)|^2$. Fig.~\ref{fig:GPE_fig} (bottom row) shows snapshots of the state as it evolves.

\noindent\emph{Code Analysis---}In the following, we demonstrate how one can code the above physics using QuSpin. As usual, we begin by loading the necessary packages:
\lstinputlisting[firstline=1, lastline=7, firstnumber=1]{\GPcode}
Next, we define the model parameters. We distinguish between static parameters and dynamic parameters -- those involved in the time-dependent trap widening.
\lstinputlisting[firstline=9, lastline=23, firstnumber=9]{\GPcode}
In order to do time evolution, we code up the trap widening protocol from Eq.~\eqref{eq:GPE} in the function \texttt{ramp}. Since we want to make use of QuSpin's time-dependent operator features, the first argument must be the time \texttt{t}, followed by all protocol parameters. These same parameters are then explicitly stored in the variable \texttt{ramp\_args}.
\lstinputlisting[firstline=24, lastline=28, firstnumber=24]{\GPcode}
With this, we are ready to construct the single-particle Hamiltonian $H_\mathrm{sp}(t)$. The first step is to define the site-coupling lists, and the static and dynamic lists. Note that the dynamic list, which defines the harmonic potential in the single-particle Hamiltonian, contains four elements: apart from the operator string and the corresponding site-coupling list, the third and fourth elements are the time-dependent function \texttt{ramp} and its argument list \texttt{ramp\_args}. We emphasize that this order of appearance is crucial.
\lstinputlisting[firstline=30, lastline=36, firstnumber=30]{\GPcode}
To create the single-particle Hamiltonian, we choose to use the bosonic basis constructor \texttt{boson\_basis\_1d} specifying the number sector to \texttt{Nb=1} boson for the entire lattice, and a local Hilbert space of \texttt{sps=2} states per site (empty and filled). 
\lstinputlisting[firstline=37, lastline=38, firstnumber=37]{\GPcode}
Then we call the \texttt{hamiltonian} constructor to build the single-particle matrix. We can obtain the single-particle ground state without fully diagonalising this matrix, by using the sparse diagonalisation method \texttt{Hsp.eigsh()}. The \texttt{eigsh()} routine accepts the optional flags \texttt{k=1} and \texttt{'which'='SA'} which restrict the underlying Lanczos routine to find the first eigenstate starting from the bottom of the spectrum, i.e.~the ground state.	
\lstinputlisting[firstline=39, lastline=41, firstnumber=39]{\GPcode}
Having set up the Hamiltonian, the next step in the simulation is to compute the ground state of the GPE using imaginary time evolution, c.f.~Eq.~\eqref{eq:GPE_imag}. To this end, we first define the function \texttt{GPE\_imag\_time} which evaluates the RHS. It is required that the first argument for this function is (imaginary) time \texttt{tau}, followed by the state \texttt{phi}. All other arguments, such as the single-particle Hamiltonian and the interaction strength are listed last. Note that we evaluate the time-dependent Hamiltonian at \texttt{time=0}, since we are interested in finding the GPE ground state for the initial trap configuration $\kappa_i$. Similar to before, we store these optional arguments in a list which we call \texttt{GPE\_params}. 
\lstinputlisting[firstline=43, lastline=51, firstnumber=43]{\GPcode}
Any initial value problem requires us to pick an initial state. In the case of imaginary evolution, this state can often be arbitrary, but needs to be in the same symmetry-sector as the true GPE ground state. Here, we choose the ground state of the single-particle Hamiltonian for an initial state, and normalise it to one particle per site. We also define the imaginary time vector \texttt{tau}. This array has to contain sufficiently long times so that we make sure we end up in the long imaginary time limit $\tau\to\infty$, as required by Eq.~\eqref{eq:GPE_imag}. Since imaginary time evolution is not unitary, QuSpin will be normalising the vector every $\tau$-step. Thus, one also needs to make sure these steps are small enough to avoid convergence problems with the ODE solver.
\lstinputlisting[firstline=52, lastline=55, firstnumber=52]{\GPcode}
Performing imaginary time evolution is done using the \texttt{evolve()} method of the \texttt{evolution} submodule. This function accepts an initial state \texttt{phi0}, initial time \texttt{tau[0]}, and a time vector \texttt{tau} and solves any user-defined ODE, here \texttt{GPE\_imag\_time}. The parameters of the ODE are passed using the keyword argument \texttt{f\_params=GPE\_params}. To ensure the normalisation of the state at each $\tau$-step we use the flag \texttt{imag\_time=True}. Real-valued output can be specified by \texttt{real=True}. Last, we request \texttt{evolve()} to create a generator object using the keyword argument \texttt{iterate=True}. Many of the keyword arguments of \texttt{evolve()} are the same as in the \texttt{H.evolve()} method of the \texttt{hamiltonian} class: for instance, one can choose a specific SciPy solver and pass its arguments, or the solver's absolute and relative tolerance. We refer the interested reader to the documentation, cf.~App.~\ref{app:doc}.
\lstinputlisting[firstline=56, lastline=58, firstnumber=56]{\GPcode}
Looping over the generator \texttt{phi\_tau} we have access to the solution, which we display in a form of sequential snapshots:
\lstinputlisting[firstline=60, lastline=77, firstnumber=60]{\GPcode}
Last, we use our GPE ground state, to time-evolve it in \emph{real} time according to the trap widening protocol \texttt{ramp}, hard-coded into the single-particle Hamiltonian. We proceed analogously -- first we define the real-time GPE and the time vector. In defining the GPE function, we split the ODE into a time-independent static part and a time-dependent dynamic part. The single-particle Hamiltonian for the former is accessed using the \texttt{hamiltonian} attribute \texttt{Hsp.static} which returns a SciPy sparse array. We can then manually add the non-linear cubic mean-field interaction term. In order to access the time-dependent part of the Hamiltonian, and evaluate it, we loop over the attribute \texttt{Hsp.dynamic}, which is a dictionary, whose keys \texttt{fun()} are time-dependent function objects already evaluated at the passed parameter values  (here \texttt{fun(t)=ramp(t,ramp\_args)}). These functions \texttt{fun()} accept only one argument: the time value. The value corresponding to the key of the \texttt{H\_sp.dynamic} dictionary is the matrix for the operator the time-dependent functions couple to, as given in the \texttt{dynamic} list used to define the Hamiltonian, here denoted by \texttt{Hd}. In the very end, we multiply the final output vector by the Schr\"odinger $-i$, which ensures the unitarity of real-time evolution.
\lstinputlisting[firstline=79, lastline=92, firstnumber=79]{\GPcode}
To perform the real-time evolution explicitly we once again use the \texttt{evolve()} function. This time, however, since the solution of the GPE is anticipated to be complex-valued, and because we do not do imaginary time, we do not need to pass the flags \texttt{real} and \texttt{imag\_time}. Instead, we decided to show the flags for the relative and absolute tolerance of the solver. 
\lstinputlisting[firstline=93, lastline=94, firstnumber=93]{\GPcode}
Finally, we can enjoy the ``movie" displaying real-time evolution
\lstinputlisting[firstline=96, lastline=114, firstnumber=96]{\GPcode}

The complete code including the lines that produce Fig.~\ref{fig:GPE_fig} is available under:\\

\href{http://weinbe58.github.io/QuSpin/examples/example8.html}{http://weinbe58.github.io/QuSpin/examples/example8.html}\\

\subsection{Integrability Breaking and Thermalising Dynamics in the Translationally Invariant 2D Transverse-Field Ising Model}
\label{subsec:higher_spin}

This example shows how to:
\begin{itemize}
	\item define symmetries for 2D systems,
	\item use the \texttt{spin\_basis\_general} class to define custom symmetries,
	\item use the \texttt{obs\_vs\_time} routine with custom user-defined generator to calculate the expectation value of operators as a function of time,
	\item use the new functionality of the basis class to calculate the entanglement entropy.
\end{itemize}

\noindent{\emph{Physics Setup---}} In Sec.~\ref{subsec:JW}, we introduced the one-dimensional transverse-field Ising (TFI) model and showed how one can map it to an exactly-solvable quadratic fermion Hamiltonian using the Jordan-Wigner transformation. This transformation allows one to obtain an exact closed-form analytic expression for the eigenstates and eigenenergies of the Hamiltonian (at least when the system obeys translational invariance). The fact that such a transformation exists, is intrinsically related to the model being integrable. Integrable models feature an extensive amount of local integrals of motion -- conserved quantities which impose specific selection rules for the transitions between the many-body states of the system. As a result, such models thermalise in a restricted sense, i.e.~in conformity with all these integrals of motion, and the long-time behaviour of the system is captured by a Generalised Gibbs Ensemble~\cite{TD_review}. The system thermalises but thermalisation is constrained. On the other hand, non-integrable systems have a far less restricted phase space, and feature unconstrained thermalisation. Their long-time behaviour is, therefore, captured by the Gibbs Ensemble. Since thermalising dynamics appears to be intrinsically related to the lack of integrability, if detected, one can use it as an indicator for the absence of a simple, closed form expression for the eigenstates and the eigenenergies of generic quantum many-body systems.

Imagine you are given a model, and you want to determine its long-time thermalisation behaviour, from which in general you can infer information about integrability. One way to proceed is to conceive a numerical experiment as follows: if we subject the system to periodic driving at an intermediate driving frequency, this will break energy conservation. Then, in the non-integrable case, the system will keep absorbing energy from the drive until the state reaches infinite-temperature. On the contrary, in the integrable scenario unlimited energy absorption will be inhibited by the existence of additional conservation laws, and the system is likely to get stuck at some energy density. Hence, the long-time dynamics of out-of-equilibrium many-body systems can be used to extract useful information about the complexity of the underlying Hamiltonian.

Below, we consider two models with periodic boundary conditions: (i) the 1d transverse field Ising model , and (ii) the 2d transverse field Ising model on a square lattice. There is no known simple mapping to a quadratic Hamiltonian in $2d$ and, therefore, the 2d Ising model is generally considered to no longer be integrable. Here, we study thermalization through energy absorption by periodically driving the two spin systems, and demonstrate the difference in the long-time heating behavior between the two. To this end, let us define two operators
\begin{equation}
H_{zz} = -\sum_{\langle ij\rangle} S^z_iS^z_{j}, \qquad H_{x} = -\sum_{i}S^x_i
\end{equation}
where $\langle ij\rangle$ restricts the sum over nearest neighbours, and use them to construct the following piecewise periodic Hamiltonian $H(t)=H(t+T)$:
\begin{equation}
H(t)=\Bigg\{ \begin{array}{cc}
H_{zz} +AH_x,& \qquad t\in[0,T/4), \\
H_{zz} -AH_x,& \qquad t\in[T/4,3T/4),\\
H_{zz} +AH_x,& \qquad t\in[3T/4,T)
\end{array}
\end{equation}
with $T$ the period of the drive, and $A$ and $\Omega=2\pi/T$ -- the corresponding driving amplitude and frequency, respectively. The dynamics is initialised in the ferromagnetic ground state of $H_{zz}$, $|\psi_i\rangle$. As the Hamiltonian obeys translation, parity and spin-inversion symmetry at all times, we will use this to reach larger system sizes by working in the symmetry sector which contains the ground state. We are only interested in the evolution of the state at integer multiples of the driving period $T$, i.e.~stroboscopically. The state at the $\ell$-th period is conveniently obtained by successive application of the Floquet unitary $U_F$:
\begin{eqnarray}
|\psi(\ell T)\rangle = [U_F(T)]^\ell|\psi_i\rangle, \qquad U_F(T)=\mathrm e^{-iT/4(H_{zz} +AH_x)}
\mathrm e^{-iT/2(H_{zz} -AH_x)}\mathrm e^{-iT/4(H_{zz} +AH_x)}
\end{eqnarray} 

In order to see the difference in the long-time heating behaviour between $1d$ and $2d$, we measure the expectation value of $H_{zz}$ as a function of time. Let us define the relative energy $Q$ absorbed from the drive:
\begin{equation}
Q(t) =\left\langle\psi(t)\left\vert\frac{ \left(H_{zz}-E_\mathrm{min}\right)}{E_\mathrm{inf.~temp.}-E_\mathrm{min}}\right\vert\psi(t)\right\rangle=\left\langle\psi(t)\left\vert\frac{ H_{zz}-E_\mathrm{min}}{-E_\mathrm{min}}\right\vert\psi(t)\right\rangle
\end{equation}
where the last equality comes from the fact that every state in the spectrum of $H_{zz}$ has a partner of opposite energy, and thus $E_\mathrm{inf.~temp.}=0$. This quantity is normalised such that an infinite-temperature state has $Q=1$, while a zero-temperature state: $Q=0$. 

Since $H_{zz}$ is not the only available observable, one might worry that some of the information contained in the quantum state is not captured in $Q(t)$. Hence, we also look at the entanglement entropy density (which is a property of the quantum state itself)
\begin{equation}
s_\mathrm{ent}(t) = -\frac{1}{|\mathrm{A}|}\mathrm{tr}_{\mathrm{A}}\left[\rho_\mathrm{A}(t)\log\rho_\mathrm{A}(t)\right], \qquad \rho_\mathrm{A}(t) = \mathrm{tr}_{\mathrm{A^c}} |\psi(t)\rangle\langle\psi(t)|
\end{equation}
of subsystem A, defined to contain the left half of the chain, and $|\mathrm{A}|=L/2$. We denoted the reduced density matrix of subsystem A by $\rho_\mathrm{A}$, and $\mathrm{A^c}$ is the complement of A. To obtain a dimensionless quantity, we normalise the entanglement entropy by the corresponding Page value~\cite{page_93}.

\begin{figure}[t!]
	\begin{subfigure}[b]{0.496\textwidth}
		\includegraphics[width=0.9\textwidth]{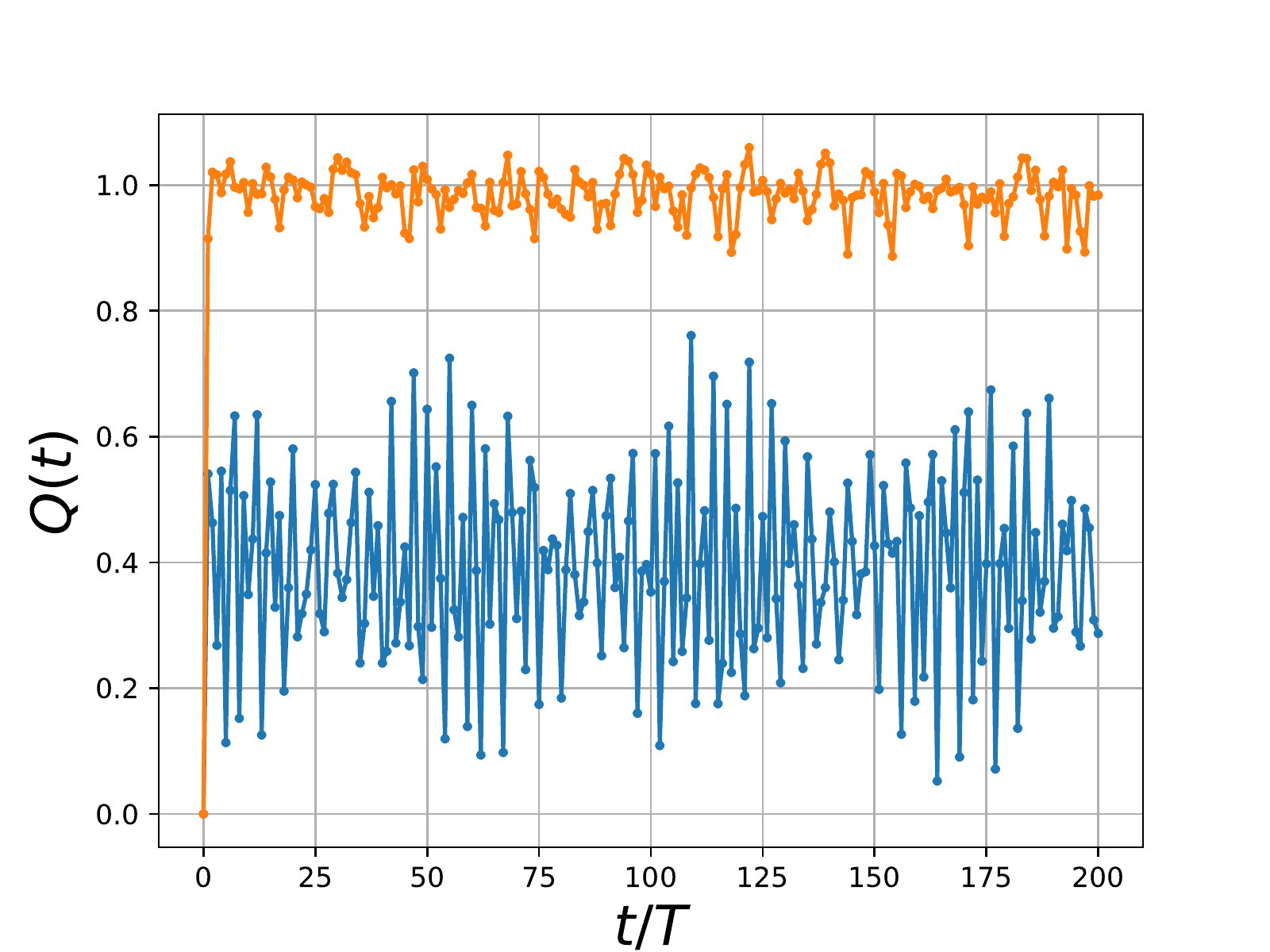}
		\caption{$Q(t)$ for $t=\ell T$}
	\end{subfigure}
	\begin{subfigure}[b]{0.496\textwidth}
		\includegraphics[width=0.9\textwidth]{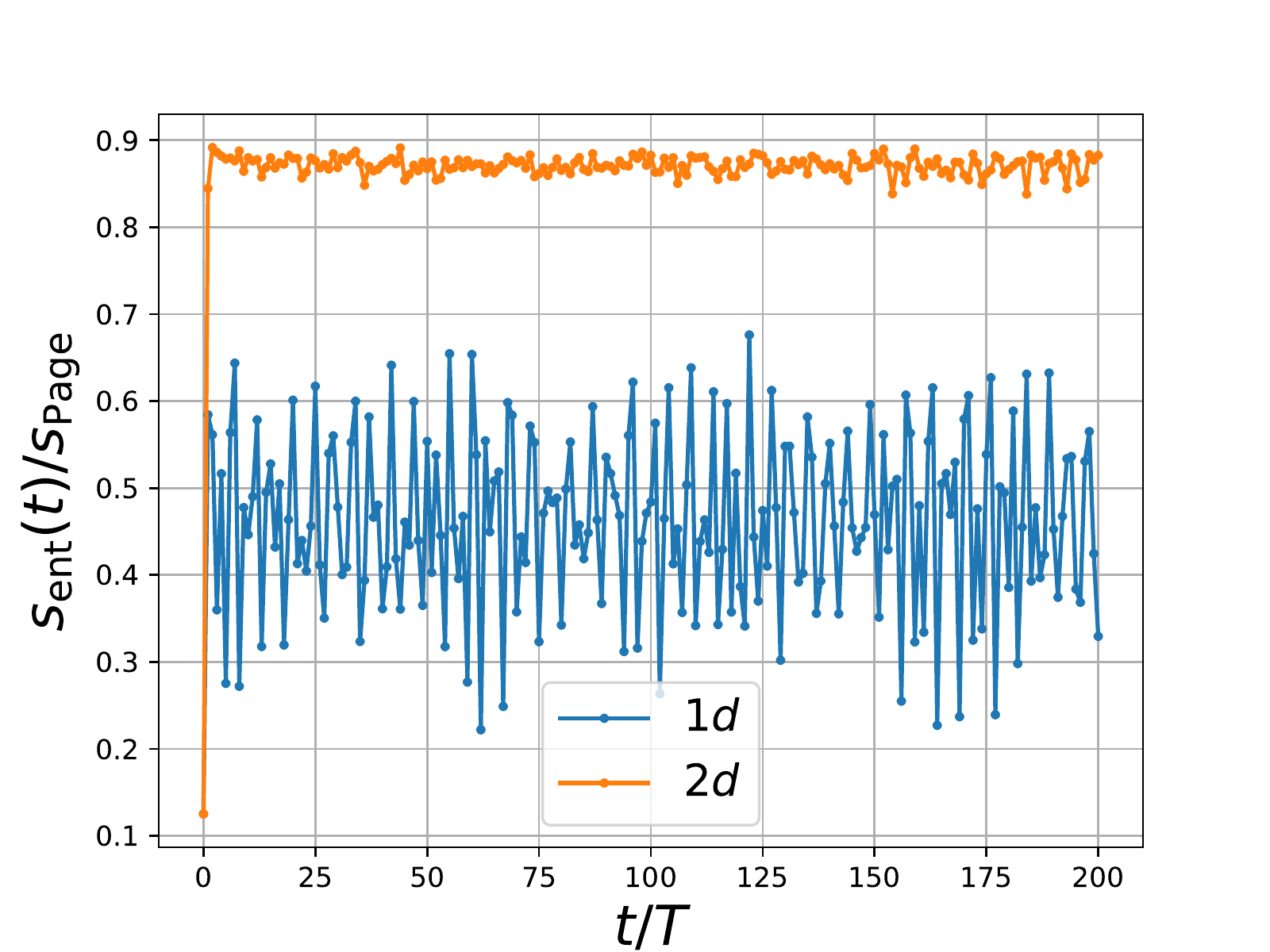}
		\caption{$s_\mathrm{ent}(t)/s_\mathrm{Page}$ for $t=\ell T$}
	\end{subfigure}
	\caption{\label{fig:example9} Comparing the dynamics of $Q(t)$ (a) and $S_\mathrm{ent}(t)$ (b) for $1d$ (orange) and $2d$ (blue) at stroboscopic times ($t=nT$). Both systems contain $N=16$ sites to make sure the many-body Hilbert spaces have roughly the same number of states. $s_\mathrm{ent}$ is normalized by the Page entropy per site. The drive frequency is $\Omega=4$.}
\end{figure}

The dynamics of the excess energy and the entanglement are shown in Fig.~\ref{fig:example9}. 

\noindent\emph{Code Analysis---}The code snippet below explains how to carry out the proposed study using QuSpin. We assume the reader has basic familiarity with QuSpin to set up simple time-dependent Hamiltonians. As is customary, we begin by loading the required Python libraries and packages:
\lstinputlisting[firstline=1, lastline=7, firstnumber=1]{\Spincode}
Then, we define the model parameters. Note that we choose all physical couplings to be unity, so the only parameters are the drive frequency and the drive amplitude, which is set equal to the driving frequency [the value of which is chosen arbitrarily]. The variable \texttt{L} denotes the linear dimension of the two spin-$1/2$ systems: the $1d$ chain has a total of \texttt{L\_1d} sites, while the $2d$ square lattice has linear dimensions \texttt{Lx} and \texttt{Ly}, and a total number of sites \texttt{N\_2d}. All variables indexed by \texttt{\_1d} and \texttt{\_2d} in the code below refer to the $1d$ and $2d$ system, respectively. 
\lstinputlisting[firstline=9, lastline=14, firstnumber=9]{\Spincode}
Before we move on to define the two bases, let us explain how to use the \texttt{spin\_basis\_general} class two manually define custom symmetries in $2d$. QuSpin can handle any unitary symmetry operation $Q$ of multiplicity $m_Q$ ($Q^{m_Q}=1$). We label its eigenvalues $\exp(-2\pi i q/m_Q)$ by an integer $q=\{0,1,\dots,m_Q-1\}$, which is used to set the symmetry block. For instance, if $Q=P$ is parity (reflection), then $q=0,1$ will correspond to the positive and negative parity blocks. If $Q=T$ is translation, then $q$ will label all momentum blocks. In the present case, the symmetries are translations and parity along the $x$ and $y$ lattice directions, and spin inversion. To this end, we first label all sites with integers, and store them in the variable \texttt{s}. One can think of the $2d$ system as a snake-like folded (or square-reshaped) \texttt{s}. With this representation each site \texttt{s} can be mapped to a \texttt{ix} and \texttt{iy} coordinate by the mapping: \texttt{s=ix+Lx*iy}. The action of any local symmetry can be programmed on this set of sites. For instance, this helps us define the action of the $x$ and $y$ translation operators \texttt{T\_x} and \texttt{T\_y}, by shifting all sites in the corresponding direction. Parity (reflection) operations \texttt{P\_x} and \texttt{P\_y} are also intuitive to define. Last, the spin inversion symmetry is uniquely defined through the mapping \texttt{s}$\;\mapsto\;$\texttt{-(1+s)} for every site \texttt{s} it acts on. In this way, the user can define different lattice geometries, and/or sublattice symmetries and QuSpin will do the hard job to reduce the Hilbert space to the corresponding symmetry sector.
\lstinputlisting[firstline=16, lastline=24, firstnumber=16]{\Spincode}
Let us define the spin bases. For the $1d$ case, we make use of the \texttt{spin\_basis\_1d} constructor. We note in passing that one can also study higher-spin systems by using the optional argument \texttt{S}, which accepts a string (integer or half-integer) to specify the spin vector size, see \href{http://weinbe58.github.io/QuSpin/index.html}{documentation}~\ref{app:doc}. Requesting symmetry blocks works as usual, by using the corresponding optional arguments. For the $2d$ spin system, we use the \texttt{spin\_basis\_general} class. This constructor allows the user to build basis objects with many user-defined symmetries, based on their action on the lattice sites. Unlike the $1d$-basis constructors, that have symmetry blocks with pre-defined variables which take integer values \texttt{kblock=0}, the general basis constructors accept user-defined block variable names which take a tuple for every symmetry requested: the first entry is the symmetry transformation itself, and the second one -- the integer which labels the required symmetry block, e.g.~\texttt{kxblock=(T\_x,0)}.
\lstinputlisting[firstline=26, lastline=32, firstnumber=26]{\Spincode} 
To set up the site-coupling lists for the two operators in the Hamiltonian we proceed in the usual manner. Using them, we can call the \texttt{hamiltonian} constructor to define the operators $-\sum_{\langle ij\rangle} S^z_{i}S^z_{j}$ and $-\sum_j S^x_j$. Here we keep the operators separate, in order to do the periodic step-drive evolution, which is why we do not need to define separate static and dynamic lists.
\lstinputlisting[firstline=34, lastline=47, firstnumber=34]{\Spincode}
In order to do time evolution, we need to define the initial state of our system. In this case, we start from the ground state of the Hamiltonian $-\sum_{\langle ij\rangle} S^z_{i}S^z_{j}$. We remind the reader that, since we work in a specific symmetry sector, this state may no longer be a product state. To this end, we employ the \texttt{eigsh()} method for sparse hermitian matrices of the \texttt{hamiltonian} class, where we explicitly specify that we are interested in getting a single (\texttt{k=1}) smallest algebraic (i.e.~ground state) eigenenergy, and the corresponding eigenstate (a.k.a.~the ground state), c.f.~\href{http://weinbe58.github.io/QuSpin/examples/example0.html}{Example 0} in Ref.~\cite{weinberg_17_quspin} for more details. 
\lstinputlisting[firstline=49, lastline=55, firstnumber=49]{\Spincode}
We are now set to do study the dynamics following the periodic step-drive. Before we go into the details, we note that QuSpin contains a build-in \texttt{Floquet} class under the \texttt{tools} module which can be useful for studying this and other periodically-driven systems, see \href{http://weinbe58.github.io/QuSpin/examples/example2.html}{Example 2} from Ref.~\cite{weinberg_17_quspin}. Here, instead, we focus on manually evolving the state. First, we define the number of periods we would like to stroboscopically evolve our system for. Stroboscopic evolution is one where all quantities are evaluated at integer multiple of the driving period. To set up a time vector, which explicitly hits all those points, we use the \texttt{Floquet\_t\_vec} class, which accepts as arguments the frequency \texttt{Omega}, the number of periods \texttt{nT}, and the number of points per period \texttt{len\_T}. The \texttt{Floquet\_t\_vec} class creates an object which has many useful attributes, including the stroboscopic times and their indices, the period, the starting point, etc. We invite the interested reader to check out the documentation for more information~\ref{app:doc}.
\lstinputlisting[firstline=57, lastline=60, firstnumber=57]{\Spincode}
Since the Hamiltonian is piece-wise constant, we can simulate the time evolution by exponentiating the separate terms. Note that, since we choose the driving phase (Floquet gauge) to yield a time-symmetric Hamiltonian, i.e.~$H(-t)=H(t)$, this results in evolving the system with the Hamiltonians $H_{zz}+AH_x$, $H_{zz}-AH_x$, $H_{zz}+AH_x$ for the durations $T/4,T/2,T/4$, respectively (think of the phase of the drive as that of a rectilinear cosine drive $\mathrm{sgn}\cos\Omega t$). To compute the matrix exponential of a static operator, we make use of the \texttt{exp\_op} class, where $\exp(z B) = $\texttt{exp\_op(B,a=z)} for some complex number $z$ and some operator $B$.
\lstinputlisting[firstline=61, lastline=65, firstnumber=61]{\Spincode}
In order to evolve the state itself, we demonstrate how to construct a user-defined generator function \texttt{evolve\_gen()}, which takes the initial state, the number of periods, and a sequence of unitaries within a period to apply them on the state. The generator character of the function means that it will not execute the loops it contains when called for the first time, but rather store information about them, and return the values one by one when prompted to do so later on. This is useful since otherwise we would have to loop over all times to evolve the state first, and then once again to compute the observables. As we can we below, the generator function allows us to get away with a single loop.
\lstinputlisting[firstline=66, lastline=75, firstnumber=66]{\Spincode}
Finally, we are ready to compute the time-dependent quantities of interest. In order to calculate the expectation $\langle\psi(t)|H_{zz}|\psi(t)\rangle$ QuSpin has a routine called \texttt{obs\_vs\_time()}. It accepts the time-dependent state \texttt{psi\_12\_t} (or its generator), the time vector \texttt{t.vals} to evaluate the observable at, and a dictionary, which contains all observables of interest (here \texttt{Hzz\_12}). The output of \texttt{obs\_vs\_time()} is a dictionary which contains the results: every observable is being parsed by a unique key (string) (here \texttt{"E"}), under which its expectation value will appear, evaluated at the requested times. Further, if one specifies the optional argument \texttt{return\_state=True}, the time-evolved state is also returned under the key \texttt{"psi\_t"}. 
\lstinputlisting[firstline=77, lastline=80, firstnumber=77]{\Spincode}
In fact, \texttt{obs\_vs\_time()} can also compute the entanglement entropy at every point of time (see \href{http://weinbe58.github.io/QuSpin/index.html}{documentation} or Sec.~\ref{subsec:BF_mixtures}). Instead, we decided to show how one can do this using the new functionality of the \texttt{basis} class. Each basis constructor comes with a function method \texttt{ent\_entropy()} which evaluates the entanglement entropy of a given state, and may return the reduced density matrix upon request. To compute the entanglement, the user needs to pass the state (here \texttt{Obs\_12\_t["psi\_t"]}), and a subsystem to define the partition for computing the entanglement. The method \texttt{ent\_entropy()} can handle vectorised calculations, and will compute the entanglement of the state for each point of time. The output is stored in a dictionary, and the entanglement entropy can be accessed with the key \texttt{"Sent\_A"}. Finally, to obtain the entanglement entropy density, we also normalise the results by the size of the subsystem of interest.
\lstinputlisting[firstline=81, lastline=83, firstnumber=81]{\Spincode}
In order get an intuition about the amount of entanglement generated in the system by the drive, we use as a reference entanglement the corresponding Page values, which is the average amount of entanglement between two subsystems of a random vectors in the many-body Hilbert space. 
\lstinputlisting[firstline=84, lastline=86, firstnumber=84]{\Spincode}

The complete code including the lines that produce Fig.~\ref{fig:example9} is available in under:\\

\href{http://weinbe58.github.io/QuSpin/examples/example9.html}{http://weinbe58.github.io/QuSpin/examples/example9.html}\\

\subsection{Out-of-Equilibrium Bose-Fermi Mixtures}
\label{subsec:BF_mixtures}

The last example in our tutorial shows how to
\begin{itemize}
	\item construct Hamiltonians for Bose-Fermi mixtures using the \texttt{tensor\_basis} class,
	\item periodically drive one subsystem (here the fermions),
	\item use new \texttt{tensor\_basis.index()} functionality to construct simple product states in the tensor basis,
	\item use \texttt{obs\_vs\_time} functionality to compute the evolution of the entanglement entropy of the bosons with the fermions.
\end{itemize}

\noindent\emph{Physics Setup---}The interest in the Bose-Fermi Hubbard (BFH) model is motivated from different areas of condensed matter and atomic physics. Studying the dressing of (interacting) fermionic atoms submerged in a superfluid Bose gas~\cite{grusdt_15}, the study of the sympathetic cooling technique~\cite{ketterle_08} to cool down spin-polarised fermions which do not interact in the $s$-wave channel, etc., are only a few of the experimental platforms for the rich physics concealed by Bose-Fermi mixtures (BFM). On the theoretical side, the BFH model is seen as a playground for the understanding of exotic phases of matter~\cite{stasyuk_15,bilitewski_15,scaramazza_16,bukov_14_BFM}, such as the coexistence of superfluid and checkerboard order, supersolid states, and the emergence of dressed compound particles. It is also a natural candidate for the search of manifestations of supersymmetry in condensed matter.

In this section, we study the generation of interspecies entanglement in a spinless Bose-Fermi mixture, caused by an external time-dependent drive. The Hamiltonian for the system reads
\begin{eqnarray}
\label{eq:H_BFM}
H(t) &=& H_\mathrm{b} + H_\mathrm{f}(t) + H_\mathrm{bf},\nonumber\\
H_\mathrm{b} &=& -J_\mathrm{b}\sum_{j}\left(b^\dagger_{j+1}b_j + \mathrm{h.c.}\right) - \frac{U_\mathrm{bb}}{2}\sum_j n^\mathrm{b}_j + \frac{U_\mathrm{bb}}{2}\sum_j n^\mathrm{b}_jn^\mathrm{b}_j,\nonumber\\
H_\mathrm{f}(t) &=& -J_\mathrm{f}\sum_{j}\left(c^\dagger_{j+1}c_j - c_{j+1}c^\dagger_j\right) + A\cos\Omega t\sum_j (-1)^j n^\mathrm{f}_j +  U_\mathrm{ff}\sum_j n^\mathrm{f}_jn^\mathrm{f}_{j+1},\nonumber\\
H_\mathrm{bf} &=& U_\mathrm{bf}\sum_j n^\mathrm{b}_jn^\mathrm{f}_j,
\end{eqnarray} 
where the operator $b^\dagger_j$ ($c^\dagger_j$) creates a boson (fermion) on site $j$, and the corresponding density is $n^\mathrm{b}_j=b^\dagger_jb_j$ ($n^\mathrm{f}_j=c^\dagger_jc_j$). The hopping matrix elements are denoted by $J_\mathrm{b}$ and $J_\mathrm{f}$, respectively. The bosons are subject to an on-site interaction of strength $U_\mathrm{bb}$, while the spin-polarised fermion-fermion interaction $U_\mathrm{ff}$ is effective on nearest-neighbouring sites. The bosonic and fermionic sectors are coupled through an on-site interspecies density-density interaction $U_\mathrm{bf}$. We assume unit filling for the bosons and half-filling for the fermions.

\begin{figure}[t!]
	\centering
	\includegraphics[width=0.5\textwidth]{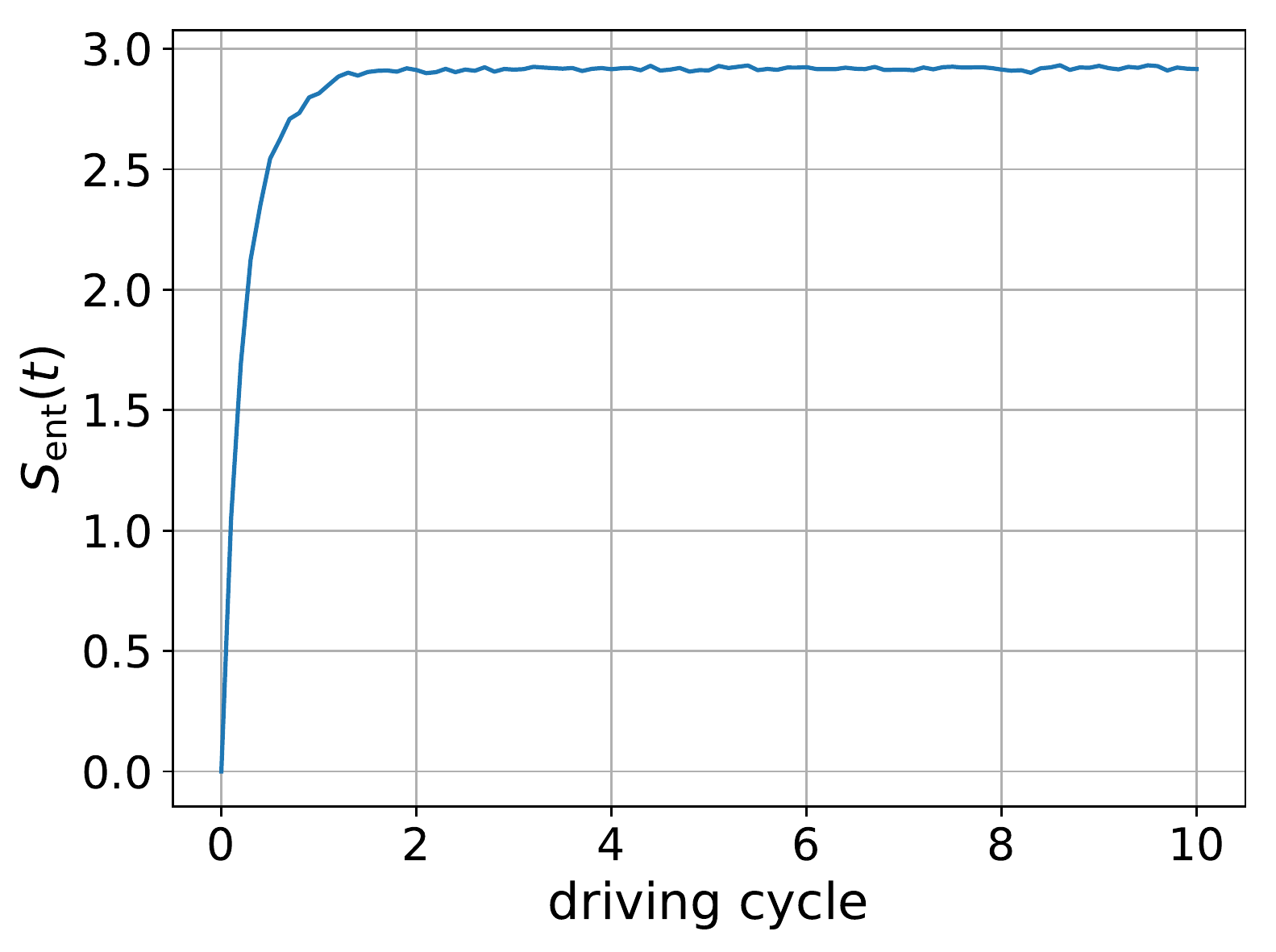}
	\caption{\label{fig:BFM} Entanglement entropy of one species as a function of time. This data was taken on a chain of length $L=6$, which agrees with the Page value~\cite{page_93} $S_\mathrm{page}\approx2.86$.}
\end{figure}

The BF mixture is initially prepared in the product state $|b\rangle|f\rangle=\prod_{j=0}^{L-1}b^\dagger_j\vert 0\rangle\prod_{j=0}^{L/2-1}c^\dagger_j\vert 0\rangle$, which is a Mott state for the bosons, and a domain wall for the fermions. A low-frequency periodic drive of amplitude $A$ and frequency $\Omega$ couples to the staggered potential in the fermions sector, and pumps energy into the system. We study the growth of the entanglement $S_\mathrm{ent}(t)$ between the two species, see Fig.~\ref{fig:BFM}.
\begin{equation}
S_\mathrm{ent}(t) = -\mathrm{tr}_\mathrm{b}\left(\rho_\mathrm{b}(t)\log\rho_\mathrm{b}(t)\right),\quad \rho_\mathrm{b}(t) = \mathrm{tr}_\mathrm{f} |\psi(t)\rangle\langle\psi(t)|,
\end{equation}  
where $\mathrm{tr}_\mathrm{b}(\cdot)$ and $\mathrm{tr}_\mathrm{f}(\cdot)$ are the traces over the boson and fermion sectors, respectively.

\noindent\emph{Code Analysis---}Let us explain how to code up the Hamiltonian for the BHM now. As always, we begin by loading the necessary modules for the simulation. New here is the \texttt{tensor\_basis} class with the help of which one can construct the basis for a tensor product Hilbert space:
\lstinputlisting[firstline=1, lastline=7, firstnumber=1]{\BFMcode}
First, we define the model parameters, and the drive:
\lstinputlisting[firstline=9, lastline=21, firstnumber=9]{\BFMcode}
Next we set up the basis, introducing the \texttt{tensor\_basis} constructor class. In its full-fledged generality, \texttt{tensor\_basis} takes $n$ basis objects which it uses to construct the matrix elements in the tensor product space: 
\begin{equation}
\mathcal{H} = \mathcal{H}_1\otimes\mathcal{H}_2\otimes\cdots\otimes\mathcal{H}_n.
\end{equation}
Here we consider the case of two Hilbert spaces, $\mathcal{H} = \mathcal{H}_1\otimes\mathcal{H}_2$, one for the bosons and one for fermions\footnote{to construct a \texttt{tensor\_basis} object in general: \texttt{t\_basis = tensor\_basis(basis\_1,basis\_2,...,basis\_n)}}. One disadvantage of the tensor basis is that it does not allow for the use of symmetries, beyond particle-conservation (magnetisation in the case of spin systems). This is because a tensor basis need not obey the symmetries of the individual bases. 
\lstinputlisting[firstline=23, lastline=27, firstnumber=23]{\BFMcode}
To create the Hamiltonian, we again use the usual form of the site-coupling lists, as if we would construct a single-species operator. Since the site-coupling lists do not yet know which operators they will refer to, this is straightforward [mind the signs for the fermionic hopping operators, though, see Sec.~\ref{subsec:SSH_model}]. We use the subscripts \texttt{b}, \texttt{f} and \texttt{bf} to designate which species this lists will refer to:
\lstinputlisting[firstline=29, lastline=40, firstnumber=29]{\BFMcode}
The new part comes in specifying the \texttt{static} and \texttt{dynamic} lists. This is where we tell QuSpin that we are dealing with two species tensored together. Notice here that the \texttt{"|"} character is used to separate the operators which belong to the boson (left side of tensor product) and fermion (right side of tensor product) Hilbert spaces in \texttt{basis}. If no operator string in present, the operator is assumed to be the identity \texttt{'I'}. The site-coupling lists, on the other hand, do not require separating the two sides of the tensor product, as it is assumed that the operator string lines up with the correct site index when the \texttt{'|'} character is removed. For instance, the only term which couples the bosons and the fermions is a density-density interaction. It is the corresponding site-coupling list, thought, that determines that it is of on-site type. This type of syntax is the same for both \texttt{static} and \texttt{dynamic} lists.
\lstinputlisting[firstline=41, lastline=54, firstnumber=41]{\BFMcode}
Computing the Hamiltonian for the BHM is done in one line:
\lstinputlisting[firstline=56, lastline=58, firstnumber=56]{\BFMcode} 
Since the basis states are coded as integers, it might be hard to find the integer corresponding to a particular Fock basis state. Therefore, the \texttt{tensor\_basis} class has a method for finding the index of a particular Fock state in the \texttt{basis}. The user just has to pass a string with zeros and ones to determine which sites are occupied and which empty. For the BHM, we choose an initial state where each site is occupied by one boson, while the fermions start in a domain wall occupying the left half of the chain. We call the index of the basis array which corresponds to the initial state \texttt{i\_0}. To create the pure initial Fock state in the full Hilbert space, we first define an array \texttt{psi\_0} which is empty, except for the position \texttt{i\_0}, where the component is set to unity by the wavefunction normalisation requirement.
\lstinputlisting[firstline=62, lastline=66, firstnumber=62]{\BFMcode} 
To compute the time evolution of the state under the Hamiltonian $H(t)$, we make use of the \texttt{Floquet\_t\_vec} class to create a time vector which hits all stroboscopic points, and further contains \texttt{len\_T=10} points per period. For the dynamics, we use the \texttt{evolve()} method of the \texttt{hamiltonian} constructor class (the details of this have been explain in Sec.~~\ref{subsec:SSH_model}). Note that we create a generator object \texttt{psi\_t}, which yields the evolved state, time step by time step.
\lstinputlisting[firstline=68, lastline=70, firstnumber=68]{\BFMcode} 
In Sec.~\ref{subsec:higher_spin}, we showed how one can use the method \texttt{basis.ent\_entropy()} to compute the time evolution of the entanglement entropy, if we have the time-evolved state. Here, we show a different way of for doing the same. We use the \texttt{measurements} function \texttt{obs\_vs\_time()}, with the user-defined generator \texttt{psi\_t} as an input. In general, \texttt{obs\_vs\_time()}, calculates the expectation of observables, see Sec.~\ref{subsec:higher_spin}. However, if we parse a non-empty dictionary, called \texttt{Sent\_args}, which contains the arguments for \texttt{basis.ent\_entropy()}, we can immediately get the result directly out of the generator. 
\lstinputlisting[firstline=71, lastline=73, firstnumber=71]{\BFMcode} 
The output of \texttt{obs\_vs\_time()} is a dictionary. The results pertaining to the calculation of entanglement
are stored under the key \texttt{"Sent\_time"}. The corresponding value itself is another dictionary, with the output of \texttt{basis.ent\_entropy()}, in which the entropy is stored under the key \texttt{"Sent\_A"}. That second dictionary can also contain other objects, such as the reduced density matrix evaluated at all time steps, if these are specified in the variable \texttt{Srdm\_args}. For more information on that, we refer the user to the documentation.
\lstinputlisting[firstline=74, lastline=75, firstnumber=74]{\BFMcode}
 
The complete code including the lines that produce Fig.~\ref{fig:BFM} is available under:\\

\href{http://weinbe58.github.io/QuSpin/examples/example10.html}{http://weinbe58.github.io/QuSpin/examples/example10.html}\\

\section{New Horizons for QuSpin}
\label{sec:outro}
As we demonstrated using various examples, QuSpin is currently capable of simulating a huge class of dynamical quantum systems. Nonetheless, we do not consider it a complete project, as we imagine adding further different functionalities, motivated by the needs of the users. Interested users can put the project on their Github watch list, which will notify them of any new releases.

Being an open project, we invite the community to fork the \href{https://github.com/weinbe58/QuSpin}{Github QuSpin repository}, and actively contribute to the development of the project! We would be more than happy to consider well-documented functions and classes, which build on QuSpin, and include them in further official releases so they can be used by the wider community. We will also consider patches which allow to combine QuSpin with other open-source packages for studying quantum physics. If you have ideas, just email us, or contact us on Github.

We would also much appreciate it, if users can properly report every single instance of a bug or malfunction in the \href{https://github.com/weinbe58/QuSpin/issues}{issues section} of the project repository, as this is an inevitable part of maturing the code.

\textbf{\textit{Parallel Capabilities.---}} Starting from version 0.3.2, QuSpin has full support for parallel computation:
\begin{itemize}
	\item OpenMP support allows to speed up calculations by parallelizing low-level loops using multiple threads. Many QuSpin functions take advantage of this within the \texttt{hamiltonian} and \texttt{basis} constructor classes, including (but not limited to) \texttt{hamiltonian.evolve()}, \texttt{hamiltonian.dot()}, \texttt{basis.Op()}, \texttt{tools.evolve.expm\_multiply\_parallel()}, etc, as well as constructing \texttt{basis} and \texttt{hamiltonian} objects themselves. It is important to mention that, in order to make use of OpenMP support, the user needs to install the \texttt{omp} version of QuSpin, which is different from the standard version. Please check \url{http://weinbe58.github.io/QuSpin/parallelization.html} for more information on how to install and use OpenMP in QuSpin.
	
	\item support of Intel's MKL linrary is inherited automatically, since QuSpin wraps some of the Numpy functionality. While this feature is not intrinsic to QuSpin per se (though it is available in QuSpin via Numpy), it can help speeding up diagonalization functions, such as \texttt{hamiltonian.eigh()} and \texttt{hamiltonian.eigsh()}, etc. Note that currently the default version of Numpy installed with anaconda has MKL support built-in. 
\end{itemize}
If you are interested in speeding up your QuSpin computation and you have multiple cores available at your disposal (e.g., on a modern desktop/laptop computer, or better yet on a computing cluster), check out \href{http://weinbe58.github.io/QuSpin/examples/example12.html#example12-label}{Example 12} online. The developers of QuSpin realize that GPU computing is becoming more important as the GPU hardware is improving rapidly compared to CPU hardware. There are a few libraries which mimic the functionality of NumPy and SciPy on GPUs. Much of that being the linear algebra operations for both sparse and dense matrices \footnote{check out the package \href{https://cupy.chainer.org/}{CuPy} for a good example of this.}. Many of the features in these libraries can be access by converting QuSpin objects to NumPy/SciPy objects which can then interact with the GPU library. However, there are two important reasons we are not directly including these libraries into QuSpin: the first reason being a lack of functionality for numerical integration of ordinary differential equations which is necessary for solving the time-dependent Schr\"odinger equation, and the second reason being that some of these libraries are are not hosted on Anaconda, which makes it impossible to use them in the packaging process for QuSpin. 

Additional features we would like to consider in future versions of QuSpin include: a basis class dedicated to single particle lattice physics allowing for more efficient and intuitive way of implementing single particle Hamiltonians, functionality for doing Lindblad and other kinds of dynamics with density matrices, finite-temperature Lanzcos methods, class for calculating correlation functions, and overall improving the efficiency of current functionality.

\section*{Acknowledgements}
We would like to thank S.~Capponi, M.~Kolodrubetz, S.~Kourtis and L.~Pollet for various stimulating discussions. Special thanks go to Alex~G.R.~Day for urging us to put a proper documentation for the package, and to W.W.~Ho for reporting some bugs in the code. We also thank all users who reported various problems on Github, and encourage them to keep doing so in the future. The authors are pleased to acknowledge that the computational work reported on in this paper was performed on the Shared Computing Cluster which is administered by \href{http://www.bu.edu/tech/support/research/}{Boston University's Research Computing Services}. The authors also acknowledge the Research Computing Services group for providing consulting support which has contributed to the results reported within this paper. We would also like to thank \href{https://github.com/}{Github} for providing the online resources to help develop and maintain this project, and the \href{http://www.sphinx-doc.org/en/stable/#}{Sphinx} project which makes it possible to maintain an easy-to-read and up-to-date \href{http://weinbe58.github.io/QuSpin/index.html}{documentation}. 

\paragraph{Funding information}
MB acknowledges support from the Emergent Phenomena in Quantum Systems initiative of the Gordon and Betty Moore Foundation, ERC synergy grant UQUAM, and the U.S. Department of Energy, Office of Science, Office of Advanced Scientific Computing Research, Quantum Algorithm Teams Program. This research was supported in part by the National Science Foundation under Grant No. NSF PHY-1748958.
\begin{appendix}
	
\section{Installation Guide in a Few Steps}
\label{app:install}
	
Detailed installation instructions can be found under:\\
	
\href{http://weinbe58.github.io/QuSpin/Installation.html}{http://weinbe58.github.io/QuSpin/Installation.html}\\

\section{Basic Use of Command Line to Run Python}
\label{app:cmd_line}

For beginners, instructions how to execute a Python code are provided under:\\

\href{http://weinbe58.github.io/QuSpin/Installation.html}{http://weinbe58.github.io/QuSpin/Installation.html}\\
	
\section{Package Documentation}
\label{app:doc}
In QuSpin quantum many-body operators are represented as matrices. The computation of these matrices are done through custom code written in Cython. Cython is an optimizing static compiler which takes code written in a syntax similar to Python, and compiles it into a highly efficient C/C\texttt{++} shared library. These libraries are then easily interfaced with Python, but can run orders of magnitude faster than pure Python code~\cite{Cython}. The matrices are stored in a sparse matrix format using the sparse matrix library of SciPy~\cite{SciPy_package}. This allows QuSpin to easily interface with mature Python packages, such as NumPy, SciPy, any many others. These packages provide reliable state-of-the-art tools for scientific computation as well as support from the Python community to regularly improve and update them~\cite{NumPy,Python_computing_1,Python_computing_2,SciPy_package}. Moreover, we have included specific functionality in QuSpin which uses NumPy and SciPy to do many desired calculations common to ED studies, while making sure the user only has to call a few NumPy or SciPy functions directly. The complete up-to-date documentation for the package is available online under:\\
	
\href{http://weinbe58.github.io/QuSpin/index.html}{http://weinbe58.github.io/QuSpin/index.html}\\
	
\section{Complete Example Codes}
\label{app:scripts}
	
The python scripts for all QuSpin examples can be downloaded under:\\

\href{http://weinbe58.github.io/QuSpin/Examples.html}{http://weinbe58.github.io/QuSpin/Examples.html}\\
	
\end{appendix}

\bibliography{qspin}

\begin{thebibliography}{100}
\providecommand{\url}[1]{\texttt{#1}}
\providecommand{\urlprefix}{URL }
\expandafter\ifx\csname urlstyle\endcsname\relax
  \providecommand{\doi}[1]{doi:\discretionary{}{}{}#1}\else
  \providecommand{\doi}{doi:\discretionary{}{}{}\begingroup
  \urlstyle{rm}\Url}\fi
\providecommand{\eprint}[2][]{\url{#2}}

\bibitem{alet05}
F.~Alet, P.~Dayal, A.~Grzesik, A.~Honecker, M.~Körner, A.~Läuchli, S.~R.
  Manmana, I.~P. McCulloch, F.~Michel, R.~M. Noack, G.~Schmid, U.~Schollwöck
  \emph{et~al.},
\newblock \emph{The alps project: Open source software for strongly correlated
  systems},
\newblock Journal of the Physical Society of Japan \textbf{74}(Suppl), 30
  (2005),
\newblock \doi{10.1143/JPSJS.74S.30},
\newblock \eprint{http://dx.doi.org/10.1143/JPSJS.74S.30}.

\bibitem{albuquerque2007}
A.~Albuquerque, F.~Alet, P.~Corboz, P.~Dayal, A.~Feiguin, S.~Fuchs, L.~Gamper,
  E.~Gull, S.~Gürtler, A.~Honecker, R.~Igarashi, M.~Körner \emph{et~al.},
\newblock \emph{The \{ALPS\} project release 1.3: Open-source software for
  strongly correlated systems},
\newblock Journal of Magnetism and Magnetic Materials \textbf{310}(2, Part 2),
  1187  (2007),
\newblock \doi{http://dx.doi.org/10.1016/j.jmmm.2006.10.304},
\newblock Proceedings of the 17th International Conference on MagnetismThe
  International Conference on Magnetism.

\bibitem{bauer11}
B.~Bauer, L.~D. Carr, H.~G. Evertz, A.~Feiguin, J.~Freire, S.~Fuchs, L.~Gamper,
  J.~Gukelberger, E.~Gull, S.~Guertler, A.~Hehn, R.~Igarashi \emph{et~al.},
\newblock \emph{The alps project release 2.0: open source software for strongly
  correlated systems},
\newblock Journal of Statistical Mechanics: Theory and Experiment
  \textbf{2011}(05), P05001 (2011).

\bibitem{dolfi14}
M.~Dolfi, B.~Bauer, S.~Keller, A.~Kosenkov, T.~Ewart, A.~Kantian, T.~Giamarchi
  and M.~Troyer,
\newblock \emph{Matrix product state applications for the \{ALPS\} project},
\newblock Computer Physics Communications \textbf{185}(12), 3430  (2014),
\newblock \doi{http://dx.doi.org/10.1016/j.cpc.2014.08.019}.

\bibitem{ITensor}
E.~M. Stoudenmire, S.~R. White \emph{et~al.},
\newblock \emph{{ITensor}: C\texttt{++} library for implementing tensor product
  wavefunction calculations} (2011--).

\bibitem{TNT}
D.~J. S.~Al-Assam, S. R.~Clark and {TNT Development team},
\newblock \emph{Tensor network theory library, beta version 1.2.0} (2016--).

\bibitem{johansson2012}
J.~Johansson, P.~Nation and F.~Nori,
\newblock \emph{Qutip: An open-source python framework for the dynamics of open
  quantum systems},
\newblock Computer Physics Communications \textbf{183}(8), 1760  (2012),
\newblock \doi{http://dx.doi.org/10.1016/j.cpc.2012.02.021}.

\bibitem{johansson2013}
J.~Johansson, P.~Nation and F.~Nori,
\newblock \emph{Qutip 2: A python framework for the dynamics of open quantum
  systems},
\newblock Computer Physics Communications \textbf{184}(4), 1234  (2013),
\newblock \doi{http://dx.doi.org/10.1016/j.cpc.2012.11.019}.

\bibitem{wright_13}
J.~G. Wright and B.~S. Shastry,
\newblock \emph{Diracq: A quantum many-body physics package},
\newblock arXiv preprint arXiv:1301.4494  (2013).

\bibitem{kramer_17}
S.~Kr{\"a}mer, D.~Plankensteiner, L.~Ostermann and H.~Ritsch,
\newblock \emph{Quantumoptics. jl: A julia framework for simulating open
  quantum systems},
\newblock Computer Physics Communications \textbf{227}, 109 (2018).

\bibitem{gegg_17}
M.~Gegg and M.~Richter,
\newblock \emph{Psiquasp--a library for efficient computation of symmetric open
  quantum systems},
\newblock Scientific reports \textbf{7}(1), 16304 (2017).

\bibitem{young_17}
L.~E. Young-S, P.~Muruganandam, S.~K. Adhikari, V.~Lon{\v{c}}ar,
  D.~Vudragovi{\'c} and A.~Bala{\v{z}},
\newblock \emph{Openmp gnu and intel fortran programs for solving the
  time-dependent gross--pitaevskii equation},
\newblock Computer Physics Communications \textbf{220}, 503 (2017).

\bibitem{al_17}
S.~Al-Assam, S.~Clark and D.~Jaksch,
\newblock \emph{The tensor network theory library},
\newblock J. Stat. Mech p. 093102 (2017).

\bibitem{weinberg_17_quspin}
P.~Weinberg and M.~Bukov,
\newblock \emph{{QuSpin: a Python Package for Dynamics and Exact
  Diagonalisation of Quantum Many Body Systems part I: spin chains}},
\newblock SciPost Phys. \textbf{2}, 003 (2017),
\newblock \doi{10.21468/SciPostPhys.2.1.003}.

\bibitem{pollet_12}
L.~Pollet,
\newblock \emph{Recent developments in quantum monte carlo simulations with
  applications for cold gases},
\newblock Reports on Progress in Physics \textbf{75}(9), 094501 (2012).

\bibitem{foulkes_01}
W.~M.~C. Foulkes, L.~Mitas, R.~J. Needs and G.~Rajagopal,
\newblock \emph{Quantum monte carlo simulations of solids},
\newblock Rev. Mod. Phys. \textbf{73}, 33 (2001),
\newblock \doi{10.1103/RevModPhys.73.33}.

\bibitem{acioli_97}
P.~H. Acioli,
\newblock \emph{Review of quantum monte carlo methods and their applications},
\newblock Journal of Molecular Structure: THEOCHEM \textbf{394}(2), 75  (1997),
\newblock \doi{http://dx.doi.org/10.1016/S0166-1280(96)04821-X}.

\bibitem{daley_04}
A.~J. Daley, C.~Kollath, U.~Schollwöck and G.~Vidal,
\newblock \emph{Time-dependent density-matrix renormalization-group using
  adaptive effective hilbert spaces},
\newblock Journal of Statistical Mechanics: Theory and Experiment
  \textbf{2004}(04), P04005 (2004).

\bibitem{schollwock_05}
U.~Schollw\"ock,
\newblock \emph{The density-matrix renormalization group},
\newblock Rev. Mod. Phys. \textbf{77}, 259 (2005),
\newblock \doi{10.1103/RevModPhys.77.259}.

\bibitem{schollwock_11}
U.~Schollwöck,
\newblock \emph{The density-matrix renormalization group in the age of matrix
  product states},
\newblock Annals of Physics \textbf{326}(1), 96  (2011),
\newblock \doi{http://dx.doi.org/10.1016/j.aop.2010.09.012},
\newblock January 2011 Special Issue.

\bibitem{georges_96}
A.~Georges, G.~Kotliar, W.~Krauth and M.~J. Rozenberg,
\newblock \emph{Dynamical mean-field theory of strongly correlated fermion
  systems and the limit of infinite dimensions},
\newblock Rev. Mod. Phys. \textbf{68}, 13 (1996),
\newblock \doi{10.1103/RevModPhys.68.13}.

\bibitem{kotliar_06}
G.~Kotliar, S.~Y. Savrasov, K.~Haule, V.~S. Oudovenko, O.~Parcollet and C.~A.
  Marianetti,
\newblock \emph{Electronic structure calculations with dynamical mean-field
  theory},
\newblock Rev. Mod. Phys. \textbf{78}, 865 (2006),
\newblock \doi{10.1103/RevModPhys.78.865}.

\bibitem{aoki_14}
H.~Aoki, N.~Tsuji, M.~Eckstein, M.~Kollar, T.~Oka and P.~Werner,
\newblock \emph{Nonequilibrium dynamical mean-field theory and its
  applications},
\newblock Rev. Mod. Phys. \textbf{86}, 779 (2014),
\newblock \doi{10.1103/RevModPhys.86.779}.

\bibitem{altman2015universal}
E.~Altman and R.~Vosk,
\newblock \emph{Universal dynamics and renormalization in many-body-localized
  systems},
\newblock Annu. Rev. Condens. Matter Phys. \textbf{6}(1), 383 (2015).

\bibitem{nandkishore_15}
R.~Nandkishore and D.~A. Huse,
\newblock \emph{Many-body localization and thermalization in quantum
  statistical mechanics},
\newblock Annual Review of Condensed Matter Physics \textbf{6}(1), 15 (2015),
\newblock \doi{10.1146/annurev-conmatphys-031214-014726},
\newblock \eprint{http://dx.doi.org/10.1146/annurev-conmatphys-031214-014726}.

\bibitem{abanin_17}
D.~A. Abanin and Z.~Papi{\'c},
\newblock \emph{Recent progress in many-body localization},
\newblock Annalen der Physik \textbf{529}(7) (2017).

\bibitem{thomson_18}
S.~J. Thomson and M.~Schir\'o,
\newblock \emph{Time evolution of many-body localized systems with the flow
  equation approach},
\newblock Phys. Rev. B \textbf{97}, 060201 (2018),
\newblock \doi{10.1103/PhysRevB.97.060201}.

\bibitem{rubio2018probing}
A.~Rubio-Abadal, J.-y. Choi, J.~Zeiher, S.~Hollerith, J.~Rui, I.~Bloch and
  C.~Gross,
\newblock \emph{Probing many-body localization in the presence of a quantum
  bath},
\newblock arXiv preprint arXiv:1805.00056  (2018).

\bibitem{TD_review}
L.~D'Alessio, Y.~Kafri, A.~Polkovnikov and M.~Rigol,
\newblock \emph{From quantum chaos and eigenstate thermalization to statistical
  mechanics and thermodynamics},
\newblock Advances in Physics \textbf{65}, 239 (2016),
\newblock \doi{10.1080/00018732.2016.1198134},
\newblock \eprint{http://dx.doi.org/10.1080/00018732.2016.1198134}.

\bibitem{nation2019ergodicity}
C.~Nation and D.~Porras,
\newblock \emph{Ergodicity probes: using time-fluctuations to measure the
  hilbert space dimension},
\newblock arXiv preprint arXiv:1906.06206  (2019).

\bibitem{duval2019quantum}
C.~Duval and M.~Kastner,
\newblock \emph{Quantum kinetic perturbation theory for near-integrable spin
  chains with weak long-range interactions},
\newblock arXiv preprint arXiv:1903.05401  (2019).

\bibitem{bentsen2019treelike}
G.~Bentsen, T.~Hashizume, A.~S. Buyskikh, E.~J. Davis, A.~J. Daley, S.~S.
  Gubser and M.~Schleier-Smith,
\newblock \emph{Treelike interactions and fast scrambling with cold atoms},
\newblock arXiv preprint arXiv:1905.11430  (2019).

\bibitem{polkovnikov_11}
A.~Polkovnikov, K.~Sengupta, A.~Silva and M.~Vengalattore,
\newblock Rev. Mod. Phys. \textbf{83}, 863 (2011).

\bibitem{goldman_14}
N.~Goldman and J.~Dalibard,
\newblock \emph{Periodically driven quantum systems: Effective hamiltonians and
  engineered gauge fields},
\newblock Phys. Rev. X \textbf{4}, 031027 (2014),
\newblock \doi{10.1103/PhysRevX.4.031027}.

\bibitem{bukov_review}
M.~Bukov, L.~D'Alessio and A.~Polkovnikov,
\newblock \emph{Universal high-frequency behavior of periodically driven
  systems: from dynamical stabilization to floquet engineering},
\newblock Advances in Physics \textbf{64}(2), 139 (2015),
\newblock \doi{10.1080/00018732.2015.1055918},
\newblock \eprint{http://dx.doi.org/10.1080/00018732.2015.1055918}.

\bibitem{eckardt_17}
A.~Eckardt,
\newblock Rev. Mod. Phys. \textbf{89}, 011004 (2017).

\bibitem{claeys2017breaking}
P.~W. Claeys and J.-S. Caux,
\newblock \emph{Breaking the integrability of the heisenberg model through
  periodic driving},
\newblock arXiv preprint arXiv:1708.07324  (2017).

\bibitem{claeys2017spin}
P.~W. Claeys, S.~De~Baerdemacker, O.~E. Araby and J.-S. Caux,
\newblock \emph{Spin polarization through floquet resonances in a driven
  central spin model},
\newblock Phys. Rev. Lett. \textbf{121}, 080401 (2018),
\newblock \doi{10.1103/PhysRevLett.121.080401}.

\bibitem{vajna2017replica}
S.~Vajna, K.~Klobas, T.~c.~v. Prosen and A.~Polkovnikov,
\newblock \emph{Replica resummation of the baker-campbell-hausdorff series},
\newblock Phys. Rev. Lett. \textbf{120}, 200607 (2018),
\newblock \doi{10.1103/PhysRevLett.120.200607}.

\bibitem{weinberg_FAPT}
P.~Weinberg, M.~Bukov, L.~D’Alessio, A.~Polkovnikov, S.~Vajna and
  M.~Kolodrubetz,
\newblock \emph{Adiabatic perturbation theory and geometry of
  periodically-driven systems},
\newblock Physics Reports \textbf{688}, 1 (2017).

\bibitem{howell2018frequency}
O.~Howell, P.~Weinberg, D.~Sels, A.~Polkovnikov and M.~Bukov,
\newblock \emph{Asymptotic prethermalization in periodically driven classical
  spin chains},
\newblock Phys. Rev. Lett. \textbf{122}, 010602 (2019),
\newblock \doi{10.1103/PhysRevLett.122.010602}.

\bibitem{vogl2019flow}
M.~Vogl, P.~Laurell, A.~D. Barr and G.~A. Fiete,
\newblock \emph{Flow equation approach to periodically driven quantum systems},
\newblock Phys. Rev. X \textbf{9}, 021037 (2019),
\newblock \doi{10.1103/PhysRevX.9.021037}.

\bibitem{yang2019quantum}
D.~Yang, A.~Grankin, L.~M. Sieberer, D.~V. Vasilyev and P.~Zoller,
\newblock \emph{Quantum non-demolition measurement of a many-body hamiltonian},
\newblock arXiv preprint arXiv:1905.06444  (2019).

\bibitem{mzaouali2019long}
Z.~Mzaouali and M.~El~Baz,
\newblock \emph{Long range quantum coherence, quantum \& classical correlations
  in heisenberg xx chain},
\newblock Physica A: Statistical Mechanics and its Applications \textbf{518},
  119 (2019).

\bibitem{niu_01}
B.~Wu and Q.~Niu,
\newblock \emph{Landau and dynamical instabilities of the superflow of
  bose-einstein condensates in optical lattices},
\newblock Phys. Rev. A \textbf{64}, 061603 (2001),
\newblock \doi{10.1103/PhysRevA.64.061603}.

\bibitem{creffield_09}
C.~E. Creffield,
\newblock Phys. Rev. A \textbf{79}, 063612 (2009).

\bibitem{bukov_12}
M.~Bukov and M.~Heyl,
\newblock \emph{Parametric instability in periodically driven luttinger
  liquids},
\newblock Phys. Rev. B \textbf{86}, 054304 (2012),
\newblock \doi{10.1103/PhysRevB.86.054304}.

\bibitem{citro2015dynamical}
R.~Citro, E.~G. Dalla~Torre, L.~D’Alessio, A.~Polkovnikov, M.~Babadi, T.~Oka
  and E.~Demler,
\newblock \emph{Dynamical stability of a many-body kapitza pendulum},
\newblock Annals of Physics \textbf{360}, 694 (2015).

\bibitem{bukov_17RL}
M.~Bukov, A.~G.~R. Day, D.~Sels, P.~Weinberg, A.~Polkovnikov and P.~Mehta,
\newblock \emph{Reinforcement learning in different phases of quantum control},
\newblock Phys. Rev. X \textbf{8}, 031086 (2018),
\newblock \doi{10.1103/PhysRevX.8.031086}.

\bibitem{lellouch_17}
S.~Lellouch, M.~Bukov, E.~Demler and N.~Goldman,
\newblock \emph{Parametric instability rates in periodically driven band
  systems},
\newblock Phys. Rev. X \textbf{7}, 021015 (2017),
\newblock \doi{10.1103/PhysRevX.7.021015}.

\bibitem{naeger2018parametric}
J.~N\"ager, K.~Wintersperger, M.~Bukov, S.~Lellouch, E.~Demler, U.~Schneider,
  I.~Bloch, N.~Goldman and M.~Aidelsburger,
\newblock \emph{Parametric instabilities of interacting bosons in
  periodically-driven 1d optical lattices},
\newblock arXiv preprint arXiv:1808.07462  (2018).

\bibitem{boulier2018parametric}
T.~Boulier, J.~Maslek, M.~Bukov, C.~Bracamontes, E.~Magnan, S.~Lellouch,
  E.~Demler, N.~Goldman and J.~V. Porto,
\newblock \emph{Parametric heating in a 2d periodically driven bosonic system:
  Beyond the weakly interacting regime},
\newblock Phys. Rev. X \textbf{9}, 011047 (2019),
\newblock \doi{10.1103/PhysRevX.9.011047}.

\bibitem{kolodrubetz_17}
M.~Kolodrubetz, D.~Sels, P.~Mehta and A.~Polkovnikov,
\newblock Physics Reports \textbf{697}, 1 (2017).

\bibitem{delcampo_13}
A.~del Campo,
\newblock \emph{Shortcuts to adiabaticity by counterdiabatic driving},
\newblock Phys. Rev. Lett. \textbf{111}, 100502 (2013),
\newblock \doi{10.1103/PhysRevLett.111.100502}.

\bibitem{sels_16}
D.~Sels and A.~Polkovnikov,
\newblock \emph{Minimizing irreversible losses in quantum systems by local
  counterdiabatic driving},
\newblock Proceedings of the National Academy of Sciences \textbf{114}(20),
  E3909 (2017),
\newblock \doi{10.1073/pnas.1619826114}.

\bibitem{bukov_GSL}
M.~Bukov, D.~Sels and A.~Polkovnikov,
\newblock \emph{Geometric speed limit of accessible many-body state
  preparation},
\newblock Phys. Rev. X \textbf{9}, 011034 (2019),
\newblock \doi{10.1103/PhysRevX.9.011034}.

\bibitem{claeys2019floquet}
P.~W. Claeys, M.~Pandey, D.~Sels and A.~Polkovnikov,
\newblock \emph{Floquet-engineering counterdiabatic protocols in quantum
  many-body systems},
\newblock arXiv preprint arXiv:1904.03209  (2019).

\bibitem{heyl2018dynamical}
M.~Heyl,
\newblock \emph{Dynamical quantum phase transitions: a review},
\newblock Reports on Progress in Physics \textbf{81}(5), 054001 (2018).

\bibitem{de2018stochastic}
S.~De~Nicola, B.~Doyon and M.~J. Bhaseen,
\newblock \emph{Stochastic approach to non-equilibrium quantum spin systems},
\newblock Journal of Physics A: Mathematical and Theoretical  (2018).

\bibitem{zache2019dynamical}
T.~V. Zache, N.~Mueller, J.~T. Schneider, F.~Jendrzejewski, J.~Berges and
  P.~Hauke,
\newblock \emph{Dynamical topological transitions in the massive schwinger
  model with a $\ensuremath{\theta}$ term},
\newblock Phys. Rev. Lett. \textbf{122}, 050403 (2019),
\newblock \doi{10.1103/PhysRevLett.122.050403}.

\bibitem{mehta2019high}
P.~Mehta, M.~Bukov, C.-H. Wang, A.~G. Day, C.~Richardson, C.~K. Fisher and
  D.~J. Schwab,
\newblock \emph{A high-bias, low-variance introduction to machine learning for
  physicists},
\newblock Physics Reports  (2019).

\bibitem{dunjko_17}
V.~Dunjko and H.~J. Briegel,
\newblock \emph{Machine learning \& artificial intelligence in the quantum
  domain: a review of recent progress},
\newblock Reports on Progress in Physics \textbf{81}(7), 074001 (2018).

\bibitem{carleo2019machine}
G.~Carleo, I.~Cirac, K.~Cranmer, L.~Daudet, M.~Schuld, N.~Tishby,
  L.~Vogt-Maranto and L.~Zdeborov{\'a},
\newblock \emph{Machine learning and the physical sciences},
\newblock arXiv preprint arXiv:1903.10563  (2019).

\bibitem{bukov2018reinforcement}
M.~Bukov,
\newblock \emph{Reinforcement learning for autonomous preparation of
  floquet-engineered states: Inverting the quantum kapitza oscillator},
\newblock Phys. Rev. B \textbf{98}, 224305 (2018),
\newblock \doi{10.1103/PhysRevB.98.224305}.

\bibitem{torlai2019integrating}
G.~Torlai, B.~Timar, E.~P. van Nieuwenburg, H.~Levine, A.~Omran, A.~Keesling,
  H.~Bernien, M.~Greiner, V.~Vuleti{\'c}, M.~D. Lukin \emph{et~al.},
\newblock \emph{Integrating neural networks with a quantum simulator for state
  reconstruction},
\newblock arXiv preprint arXiv:1904.08441  (2019).

\bibitem{fabiani2019investigating}
G.~Fabiani and J.~Mentink,
\newblock \emph{Investigating ultrafast quantum magnetism with machine
  learning},
\newblock arXiv preprint arXiv:1903.08482  (2019).

\bibitem{glaser_15}
S.~J. Glaser, U.~Boscain, T.~Calarco, C.~P. Koch, W.~K{\"o}ckenberger,
  R.~Kosloff, I.~Kuprov, B.~Luy, S.~Schirmer, T.~Schulte-Herbr{\"u}ggen
  \emph{et~al.},
\newblock \emph{Training schr{\"o}dinger’s cat: quantum optimal control},
\newblock The European Physical Journal D \textbf{69}(12), 279 (2015).

\bibitem{bukov_17symmbreak}
M.~Bukov, A.~G.~R. Day, P.~Weinberg, A.~Polkovnikov, P.~Mehta and D.~Sels,
\newblock \emph{Broken symmetry in a two-qubit quantum control landscape},
\newblock Phys. Rev. A \textbf{97}, 052114 (2018),
\newblock \doi{10.1103/PhysRevA.97.052114}.

\bibitem{day2018glassy}
A.~G.~R. Day, M.~Bukov, P.~Weinberg, P.~Mehta and D.~Sels,
\newblock \emph{Glassy phase of optimal quantum control},
\newblock Phys. Rev. Lett. \textbf{122}, 020601 (2019),
\newblock \doi{10.1103/PhysRevLett.122.020601}.

\bibitem{patil2019obstacles}
P.~Patil, S.~Kourtis, C.~Chamon, E.~R. Mucciolo and A.~E. Ruckenstein,
\newblock \emph{Obstacles to quantum annealing in a planar embedding of
  xorsat},
\newblock arXiv preprint arXiv:1904.01860  (2019).

\bibitem{kubiczek2019exact}
P.~Kubiczek, A.~N. Rubtsov and A.~I. Lichtenstein,
\newblock \emph{Exact real-time dynamics of single-impurity anderson model from
  a single-spin hybridization-expansion},
\newblock arXiv preprint arXiv:1904.12582  (2019).

\bibitem{sachdev_book}
S.~Sachdev,
\newblock \emph{Quantum Phase Transitions},
\newblock Cambridge University Press, New York (1997).

\bibitem{dziarmaga_10}
J.~Dziarmaga,
\newblock \emph{Dynamics of a quantum phase transition and relaxation to a
  steady state},
\newblock Advances in Physics \textbf{59}(6), 1063 (2010).

\bibitem{pfeuty_79}
P.~Pfeuty,
\newblock \emph{An exact result for the 1d random ising model in a transverse
  field},
\newblock Physics Letters A \textbf{72}(3), 245  (1979),
\newblock \doi{https://doi.org/10.1016/0375-9601(79)90017-3}.

\bibitem{SSH_1}
W.~P. Su, J.~R. Schrieffer and A.~J. Heeger,
\newblock \emph{Solitons in polyacetylene},
\newblock Phys. Rev. Lett. \textbf{42}, 1698 (1979),
\newblock \doi{10.1103/PhysRevLett.42.1698}.

\bibitem{SSH_2}
W.~P. Su, J.~R. Schrieffer and A.~J. Heeger,
\newblock \emph{Soliton excitations in polyacetylene},
\newblock Phys. Rev. B \textbf{22}, 2099 (1980),
\newblock \doi{10.1103/PhysRevB.22.2099}.

\bibitem{SSH_2_E}
W.~P. Su, J.~R. Schrieffer and A.~J. Heeger,
\newblock \emph{Erratum: Soliton excitations in polyacetylene},
\newblock Phys. Rev. B \textbf{28}, 1138 (1983),
\newblock \doi{10.1103/PhysRevB.28.1138}.

\bibitem{asboth2016short}
J.~K. Asb{\'o}th, L.~Oroszl{\'a}ny and A.~P{\'a}lyi,
\newblock \emph{A short course on topological insulators},
\newblock Lecture Notes in Physics \textbf{919} (2016).

\bibitem{khemani_16}
V.~Khemani, F.~Pollmann and S.~L. Sondhi,
\newblock \emph{Obtaining highly excited eigenstates of many-body localized
  hamiltonians by the density matrix renormalization group approach},
\newblock Phys. Rev. Lett. \textbf{116}, 247204 (2016),
\newblock \doi{10.1103/PhysRevLett.116.247204}.

\bibitem{khemani_17}
V.~Khemani, S.~P. Lim, D.~N. Sheng and D.~A. Huse,
\newblock \emph{Critical properties of the many-body localization transition},
\newblock Phys. Rev. X \textbf{7}, 021013 (2017),
\newblock \doi{10.1103/PhysRevX.7.021013}.

\bibitem{ponte_17}
P.~Ponte, C.~Laumann, D.~A. Huse and A.~Chandran,
\newblock \emph{Thermal inclusions: how one spin can destroy a many-body
  localized phase},
\newblock Philosophical Transactions of the Royal Society A: Mathematical,
  Physical and Engineering Sciences \textbf{375}(2108), 20160428 (2017).

\bibitem{kondov_15}
S.~S. Kondov, W.~R. McGehee, W.~Xu and B.~DeMarco,
\newblock \emph{Disorder-induced localization in a strongly correlated atomic
  hubbard gas},
\newblock Phys. Rev. Lett. \textbf{114}, 083002 (2015),
\newblock \doi{10.1103/PhysRevLett.114.083002}.

\bibitem{Schreiber15}
M.~Schreiber, S.~S. Hodgman, P.~Bordia, H.~P. L{\"u}schen, M.~H. Fischer,
  R.~Vosk, E.~Altman, U.~Schneider and I.~Bloch,
\newblock \emph{Observation of many-body localization of interacting fermions
  in a quasirandom optical lattice},
\newblock Science \textbf{349}(6250), 842 (2015),
\newblock \doi{10.1126/science.aaa7432},
\newblock \eprint{http://science.sciencemag.org/content/349/6250/842.full.pdf}.

\bibitem{ovadia_15}
M.~Ovadia, D.~Kalok, I.~Tamir, S.~Mitra, B.~Sac\'ep\'e and D.~Shahar,
\newblock \emph{Evidence for a finite-temperature insulator},
\newblock Scientific Reports \textbf{5}, 13503 (2015),
\newblock \doi{10.1038/srep13503}.

\bibitem{bordia_16}
P.~Bordia, H.~P. L\"uschen, S.~S. Hodgman, M.~Schreiber, I.~Bloch and
  U.~Schneider,
\newblock \emph{Coupling identical one-dimensional many-body localized
  systems},
\newblock Phys. Rev. Lett. \textbf{116}, 140401 (2016),
\newblock \doi{10.1103/PhysRevLett.116.140401}.

\bibitem{smith_16}
J.~Smith, A.~Lee, P.~Richerme, B.~Neyenhuis, P.~W. Hess, P.~Hauke, M.~Heyl,
  D.~A. Huse and C.~Monroe,
\newblock \emph{Many-body localization in a quantum simulator with programmable
  random disorder},
\newblock Nature Physics \textbf{12}, 907–911 (2016),
\newblock \doi{10.1038/nphys3783}.

\bibitem{choi_16}
J.-y. Choi, S.~Hild, J.~Zeiher, P.~Schau{\ss}, A.~Rubio-Abadal, T.~Yefsah,
  V.~Khemani, D.~A. Huse, I.~Bloch and C.~Gross,
\newblock \emph{Exploring the many-body localization transition in two
  dimensions},
\newblock Science \textbf{352}(6293), 1547 (2016),
\newblock \doi{10.1126/science.aaf8834},
\newblock
  \eprint{http://science.sciencemag.org/content/352/6293/1547.full.pdf}.

\bibitem{kucsko_16}
G.~Kucsko, S.~Choi, J.~Choi, P.~C. Maurer, H.~Zhou, R.~Landig, H.~Sumiya,
  S.~Onoda, J.~Isoya, F.~Jelezko, E.~Demler, N.~Y. Yao \emph{et~al.},
\newblock \emph{Critical thermalization of a disordered dipolar spin system in
  diamond},
\newblock Phys. Rev. Lett. \textbf{121}, 023601 (2018),
\newblock \doi{10.1103/PhysRevLett.121.023601}.

\bibitem{lueschen_17}
H.~P. L\"uschen, P.~Bordia, S.~S. Hodgman, M.~Schreiber, S.~Sarkar, A.~J.
  Daley, M.~H. Fischer, E.~Altman, I.~Bloch and U.~Schneider,
\newblock \emph{Signatures of many-body localization in a controlled open
  quantum system},
\newblock Phys. Rev. X \textbf{7}, 011034 (2017),
\newblock \doi{10.1103/PhysRevX.7.011034}.

\bibitem{bordia_17}
P.~Bordia, H.~L\"uschen, S.~Scherg, S.~Gopalakrishnan, M.~Knap, U.~Schneider
  and I.~Bloch,
\newblock \emph{Probing slow relaxation and many-body localization in
  two-dimensional quasiperiodic systems},
\newblock Phys. Rev. X \textbf{7}, 041047 (2017),
\newblock \doi{10.1103/PhysRevX.7.041047}.

\bibitem{zhang_17}
J.~Zhang, G.~Pagano, P.~W. Hess, A.~Kyprianidis, P.~Becker, H.~Kaplan, A.~V.
  Gorshkov, Z.-X. Gong and C.~Monroe,
\newblock \emph{Observation of a many-body dynamical phase transition with a
  53-qubit quantum simulator},
\newblock Nature \textbf{551}(7682), 601 (2017).

\bibitem{bloch_05}
I.~Bloch, J.~Dalibard and W.~Zwerger,
\newblock \emph{Many-body physics with ultracold gases},
\newblock Rev. Mod. Phys. \textbf{80}, 885 (2008).

\bibitem{anders10}
A.~W. Sandvik,
\newblock \emph{Computational studies of quantum spin systems},
\newblock AIP Conference Proceedings \textbf{1297}(1), 135 (2010),
\newblock \doi{http://dx.doi.org/10.1063/1.3518900}.

\bibitem{dalfovo_99}
F.~Dalfovo, S.~Giorgini, L.~P. Pitaevskii and S.~Stringari,
\newblock \emph{Theory of bose-einstein condensation in trapped gases},
\newblock Rev. Mod. Phys. \textbf{71}, 463 (1999),
\newblock \doi{10.1103/RevModPhys.71.463}.

\bibitem{sulem2007nonlinear}
C.~Sulem and P.-L. Sulem,
\newblock \emph{The nonlinear Schr{\"o}dinger equation: self-focusing and wave
  collapse}, vol. 139,
\newblock Springer Science \& Business Media (2007).

\bibitem{polkovnikov2010phase}
A.~Polkovnikov,
\newblock \emph{Phase space representation of quantum dynamics},
\newblock Annals of Physics \textbf{325}(8), 1790 (2010).

\bibitem{page_93}
D.~N. Page,
\newblock Phys. Rev. Lett. \textbf{71}, 1291 (1993).

\bibitem{grusdt_15}
F.~Grusdt and E.~Demler,
\newblock \emph{New theoretical approaches to bose polarons},
\newblock arXiv preprint arXiv:1510.04934 p. 1510.04934 (2015).

\bibitem{ketterle_08}
W.~Ketterle and M.~W. Zwierlein,
\newblock \emph{Making, probing and understanding ultracold fermi gases},
\newblock arXiv preprint arXiv:0801.2500  (2008).

\bibitem{stasyuk_15}
I.~V. Stasyuk and V.~O. Krasnov,
\newblock \emph{Phase transitions in bose-fermi-hubbard model in the heavy
  fermion limit: Hard-core boson approach},
\newblock Condens. Matter Phys. \textbf{18}, 43702 (2015).

\bibitem{bilitewski_15}
T.~Bilitewski and L.~Pollet,
\newblock \emph{Exotic superconductivity through bosons in a dynamical cluster
  approximation},
\newblock Phys. Rev. B \textbf{92}, 184505 (2015),
\newblock \doi{10.1103/PhysRevB.92.184505}.

\bibitem{scaramazza_16}
J.~A.~K. Scaramazza, L.~Ben and Y.~Hong,
\newblock \emph{Competing orders in a dipolar bose-fermi mixture on a square
  optical lattice: Mean-field perspective},
\newblock Eur. Phys. J. D \textbf{70}, 147 (2016).

\bibitem{bukov_14_BFM}
M.~Bukov and L.~Pollet,
\newblock Phys. Rev. B \textbf{89}, 094502 (2014).

\bibitem{Cython}
S.~Behnel, R.~Bradshaw, C.~Citro, L.~Dalcin, D.~S. Seljebotn and K.~Smith,
\newblock \emph{Cython: The best of both worlds},
\newblock Computing in Science \& Engineering \textbf{13}(2), 31 (2011),
\newblock \doi{http://dx.doi.org/10.1109/MCSE.2010.118}.

\bibitem{SciPy_package}
E.~Jones, T.~Oliphant, P.~Peterson \emph{et~al.},
\newblock \emph{{SciPy}: Open source scientific tools for {Python}} (2001--).

\bibitem{NumPy}
S.~v.~d. Walt, S.~C. Colbert and G.~Varoquaux,
\newblock \emph{The numpy array: A structure for efficient numerical
  computation},
\newblock Computing in Science \& Engineering \textbf{13}(2), 22 (2011),
\newblock \doi{http://dx.doi.org/10.1109/MCSE.2011.37}.

\bibitem{Python_computing_1}
T.~E. Oliphant,
\newblock \emph{Python for scientific computing},
\newblock Computing in Science \& Engineering \textbf{9}(3), 10 (2007),
\newblock \doi{http://dx.doi.org/10.1109/MCSE.2007.58}.

\bibitem{Python_computing_2}
K.~J. Millman and M.~Aivazis,
\newblock \emph{Python for scientists and engineers},
\newblock Computing in Science \& Engineering \textbf{13}(2), 9 (2011),
\newblock \doi{http://dx.doi.org/10.1109/MCSE.2011.36}.

\end{thebibliography}


\begin{thebibliography}{99}
\bibitem{1931_Bethe_ZP_71} H. A. Bethe, {\it Zur Theorie der Metalle. i. Eigenwerte und Eigenfunktionen der linearen Atomkette}, Zeit. f{\"u}r Phys. {\bf 71}, 205 (1931), \doi{10.1007\%2FBF01341708}.
\bibitem{arXiv:1108.2700} P. Ginsparg, {\it It was twenty years ago today... }, \url{http://arxiv.org/abs/1108.2700}.
\end{thebibliography}

\nolinenumbers

\end{document}